\documentclass[aps,twocolumn,superscriptaddress]{revtex4-2}
\usepackage{mathtools}
\usepackage{bm}
\usepackage{times}
\usepackage{dsfont,amsthm,amsbsy}
\usepackage{verbatim}
\usepackage{amssymb}
\usepackage{amsmath}
\usepackage{bbm}
\usepackage{graphicx}
\usepackage{epstopdf}
\usepackage{subfigure}
\usepackage{natbib}
\usepackage{epsfig}
\usepackage{xcolor}
\usepackage{amsfonts}
\usepackage{mathrsfs}
\usepackage{sidecap}
\usepackage{lipsum}
\usepackage[toc,page,title,titletoc,header]{appendix}
\usepackage[colorlinks,linkcolor=blue,citecolor=blue,anchorcolor=blue,urlcolor=blue]{hyperref}
\usepackage{hyperref}
\usepackage{resizegather}
\usepackage{tikz}
\usepackage{float}
\usepackage{mathbbol}
\usepackage[normalem]{ulem}
\usepackage{cancel}
\usepackage{upgreek}

\newcommand{\bl}{\begin{aligned}}
\newcommand{\el}{\end{aligned}}

\def\be{\begin{equation}}
\def\ee{\end{equation}}

\def\bi{\begin{itemize}}
\def\ei{\end{itemize}}
\def\bn{\begin{enumerate}}
\def\en{\end{enumerate}}
\def\bea{\begin{eqnarray}}
\def\eea{\end{eqnarray}}
\def\no{\nonumber}
\def\ba{\begin{array}}
\def\ea{\end{array}}
\def\bd{\begin{displaymath}}
\def\ed{\end{displaymath}}

\begin{document}
\title
{Quantum Correlation Dynamics Subjected to Quantum Reset-Driven Environment}


\author{R. Jafari}
\email[]{raadmehr.jafari@gmail.com}
\affiliation{Department of Physics, Institute for Advanced Studies in Basic Sciences (IASBS), Zanjan 45137-66731, Iran}
\affiliation{School of Quantum Physics and Matter, Institute for Research in Fundamental Sciences (IPM), Tehran 19538-33511, Iran}


%
\author{Ali Asadian}
\email[]{ali.asadian@iasbs.ac.ir.}
\affiliation{Department of Physics, Institute for Advanced Studies in Basic Sciences (IASBS), Zanjan 45137-66731, Iran}

\author{Mehdi Biderang}
\affiliation{Department of Physics, University of Toronto, 60 St. George Street, Toronto, Ontario, M5S 1A7, Canada}

\author{Alireza Akbari}
\email[]{alireza@bimsa.cn}
\affiliation{Beijing Institute of Mathematical Sciences and Applications (BIMSA), Huairou District, Beijing 101408,  China}

\date{\today}

\begin{abstract}
We study two central qubits interacting with a transverse-field Ising chain that serves as their environment. The environment is driven linearly in time across its quantum critical points (QCPs) and, during the evolution, is subjected to quantum reset (QR), where it is returned at random times to its initial state. We investigate how such QR of the environmental spin chain modifies the dynamics of entanglement and quantum discord between the qubits. 
Our results show that in the strong-coupling regime, entanglement and discord exhibit pronounced revivals within the interval bounded by the Ising QCPs, but these revivals diminish as the QR rate increases. In contrast, weak coupling leads to a monotonic reduction of quantum correlations. Numerically, we find that the revival peaks of concurrence decay and scale exponentially with the QR rate, while quantum discord shows no clear scaling behavior. 
In the weak-coupling regime without QR, the correlations decay monotonically as the driven field crosses the second QCP. When QR is applied, however, both entanglement and discord undergo oscillatory suppression, with the oscillation period increasing as either the QR rate or the ramp time scale is reduced.
\end{abstract}

%
\maketitle

\section{Introduction}\label{intro}
Quantum correlations (QCs) play a fundamental role in quantum information \cite{Barenco1,Pereira} and quantum computation \cite{Barenco2,Grover,Jafari:2010aa,Mishra2018}, and they are deeply connected to the problem of nonlocality in quantum mechanics \cite{Einstein,Bell}. A central challenge, however, is that no quantum system used for information processing can be fully isolated from its surroundings. Inevitably, quantum systems couple to environmental spin systems (ESS), and this interaction induces noise and decoherence that degrade quantum resources \cite{Zurek2003}.  

Understanding how QCs evolve in such open settings has therefore attracted significant attention. Over the past decade, many studies have examined QC dynamics and decoherence factors of central systems embedded in diverse environments \cite{Almeida2007,Cormick2008,Kevin2001,Gu2004}. Among these, ESS that undergo quantum phase transitions (QPTs) have been of particular interest, since the enhanced sensitivity near quantum critical points (QCPs) leads to striking dynamical signatures in central systems \cite{Quan2006,Jafari2017,Damski2011,Jafari2015,Aguilar2014,Liu2010,Yuan2007,Guo2016,Sun2007,Schliemann2002,Wang2001,Korbicz2021,Cucchietti2006,
Jafari:2025ac,Suzuki2016,Nag2012}. Investigations of this type have revealed phenomena such as sudden death of entanglement \cite{Almeida2007,Yu2004,Salles2008} and sudden changes of quantum discord \cite{Maziero2009,Xu2010}. They have also inspired strategies to protect correlations against decoherence, while simultaneously providing new insights into foundational questions such as the quantum-classical boundary \cite{Zurek2003,Blume2006,Cornelio2012}.  

In parallel, there has been growing interest in nonequilibrium processes involving stochastic resetting. In its simplest form, stochastic resetting interrupts the unitary evolution of a quantum system and returns it to its initial state at randomly distributed times \cite{Evans2011,Evans2014,Mukherjee2018,Magoni2020,Majumdar2015,Sevilla2023,Yin2023,Perfetto2021,Kulkarni2023,Das2022,Chelminiak2022}. This simple mechanism has become a paradigm across statistical physics \cite{Magoni2020}, stochastic thermodynamics \cite{Gupta2020,Fuchs2016}, biological physics \cite{Lisica2016}, and biochemistry \cite{Michaelis2015}, and has recently been extended to quantum systems \cite{Sevilla2023,Yin2023,Perfetto2021,Kulkarni2023,Das2022}. A large body of work now shows that resetting can generate nontrivial nonequilibrium stationary states, modify relaxation pathways, and induce long-range correlations \cite{Evans2014,Mukherjee2018,Magoni2020,Evans2020,Nagar2023}. In quantum settings, recent studies have demonstrated that resetting can lead to qualitatively new dynamical regimes and correlation structures \cite{Mukherjee2018,Rose2018,Perfetto2021,Magoni2020,Sevilla2023,Yin2023,Yin2024,Shushin2011,Gabriele2022,Turkeshi2022,Kulkarni2023}.

In this paper we consider two central qubits coupled to a time-driven transverse-field Ising chain that serves as their environment. The transverse field is ramped linearly so that the environment is driven across its QCPs, while stochastic quantum resetting  returns it to its initial state at randomly chosen times during the evolution. Our goal is to analyze how such quantum reset (QR) modifies the dynamics of entanglement and quantum discord of the central qubits, in both strong- and weak-coupling regimes. To our knowledge, systematic studies of how QR affects the dynamics of quantum systems are still very recent \cite{Magoni2020,Turkeshi2022,Kulkarni2023}. Moreover, the characteristic consequences of resetting on central quantities of quantum dynamics such as coherence, purity, and fidelity have not yet been explored in general settings, particularly when the time intervals between reset events follow arbitrary distributions. Addressing this problem provides an opportunity to deepen the understanding of nonequilibrium correlation dynamics and to identify possible routes for controlling QCs through reset protocols.

To the best of our knowledge, while stochastic resetting has been applied to
bulk spin chains and to non-Hermitian quasiparticles~\cite{Mukherjee2018,Magoni2020,Kulkarni2023,Turkeshi2022},
its effect on \emph{central-spin quantum correlations} has not been systematically
explored. In particular, the interplay of concurrence and quantum discord with stochastic
resetting in driven critical environments remains an open question. This work
fills this gap by focusing on how resetting modifies the reduced dynamics of two
central qubits, thereby providing new insights into quantum information resources
under nonequilibrium conditions.

Building on this motivation, we demonstrate that a QR-driven environment profoundly reshapes correlation dynamics. In the strong-coupling regime, concurrence and quantum discord exhibit near-perfect revivals within the interval bounded by the Ising QCPs, but these revivals diminish as the QR rate increases. In weak coupling, both measures decay monotonically. Our numerical analysis further reveals that the peak heights of concurrence revivals scale \emph{exponentially} with the QR rate, whereas discord shows no clear scaling. In addition, QR induces an oscillatory suppression of correlations, with the oscillation period increasing as either the QR rate or the ramp time scale is reduced. These results highlight QR as a versatile mechanism for reshaping the dynamical behavior of quantum correlations in driven many-body environments.

The remainder of the paper is organized as follows. In Sec.~\ref{model} we introduce the central-spin setup, define the driven Ising environment, and specify the reduced dynamics framework. Section~\ref{sec:nqr} analyzes the baseline (no–reset) dynamics and the associated decoherence factor. In Sec.~\ref{sec:qr} we formulate the quantum reset (QR) protocol and derive the ensemble-averaged density matrix under QR. Section~\ref{sec:num} presents numerical results for concurrence and quantum discord across coupling regimes and driving protocols. Finally, Sec.~\ref{sec:concl} summarizes our findings and discusses outlook. 


\section{Central-Spin Architecture and Driven Ising Environment}\label{model}
\begin{figure}
\includegraphics[width=\columnwidth, clip=true]{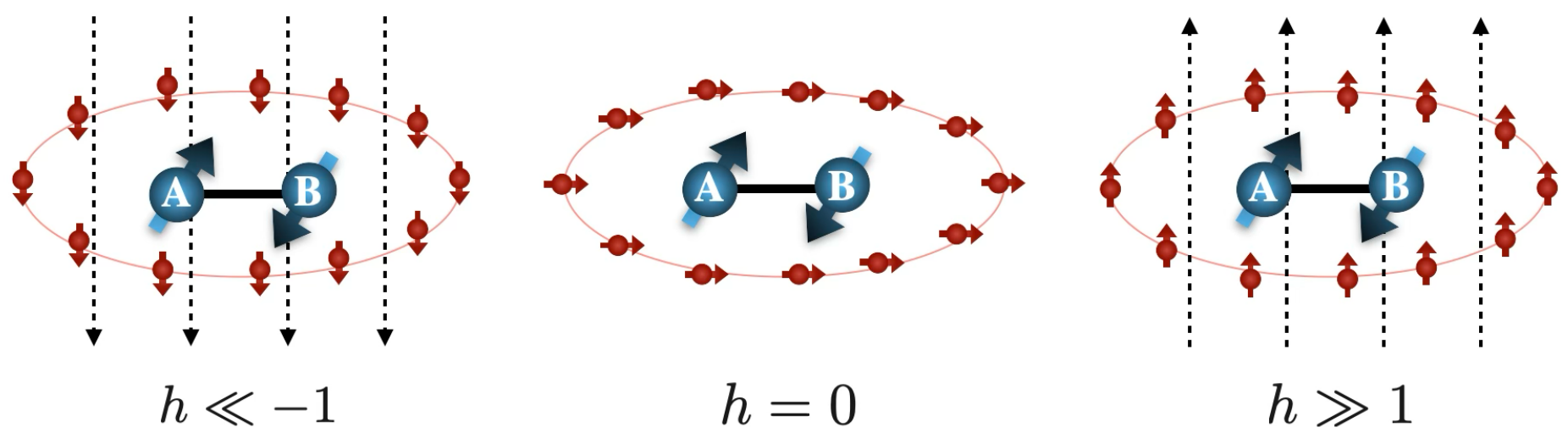}
\caption{Schematic of the driven transverse-field Ising chain that serves as the environment for the two central qubits.  
For large $|h(t)|$, the chain is in the paramagnetic phase with spins aligned along the field (left and right panels).  
For $|h(t)|<1$ (center panel), the chain enters the ferromagnetic phase, where spins align along $\pm x$, spontaneously breaking the Ising symmetry.  
The qubits couple via their $z$ components to the environment, and the field $h(t)$ is ramped across the quantum critical points at $h=\pm1$.
}
\label{fig1}
\end{figure}

In this section we investigate the dynamics of entanglement and quantum discord for two central qubits, $S_A^z=\sigma_A^z/2$ and $S_B^z=\sigma_B^z/2$, coupled to a time-dependent transverse-field Ising chain (Fig.~\ref{fig1}) \cite{Yuan2007,Guo2016,Sun2007}, where $\sigma^{\alpha}$ ($\alpha=x,y,z$) denote the Pauli matrices. Thus, the total Hamiltonian can be written as
\bea
{\cal H} = {\cal H}_{E} + {\cal H}_{AB} + {\cal H}_{I},
\eea
where, the environment is described by the time-dependent transverse-field Ising Hamiltonian  
\begin{equation}
\label{eq1}
{\cal H}_{E}(h(t))= \sum_{j=1}^N \left(\sigma _{j}^{x}\sigma_{j+1}^{x}+h(t)\sigma _{j}^{z}\right),
\end{equation}
the central qubits interact through  
\begin{equation}
\label{eq2}
{\cal H}_{AB}=S^{z}_{A}S^{z}_{B},
\end{equation}
and the coupling between qubits and environment is given by  
\begin{equation}
\label{eq3}
{\cal H}_{I}=\frac{\delta}{2} \sum_{j=1}^N\sigma _{j}^{z}(S^{z}_{A}+S^{z}_{B}).
\end{equation}
 For a static transverse field $h(t)=h$, the ground state of the environment is ferromagnetic when $|h|<1$ and paramagnetic otherwise, with the two phases separated by quantum critical points at $h_c=\pm 1$ \cite{Pfeuty1970}.
We focus on a linear ramp of the transverse field, 
$$h(t)=t/\tau,$$
 starting from an initial value 
$h_i<0$ at time $t_i$ and evolving to $h(t)$ at time $t$, where $\tau$ denotes the ramp time scale. Since $[{\cal H}_{AB},\sigma^{\alpha}_j]=0$, the operator $\tfrac{\delta}{2}(S^{z}_{A}+S^{z}_{B})$ is conserved during the evolution. As a result, the total Hamiltonian ${\cal H}$ can be expressed as
\begin{equation}
\label{eq4}
{\cal H}(t)=\sum_{\ell=1}^{4}|\phi_{\ell}\rangle\langle\phi_{\ell}|\otimes {\cal H}_{E}(h_{\ell}(t)),   
\end{equation}
where $|\phi_{\ell}\rangle$ denote the eigenstates of ${\cal H}_{AB}$, namely 
$$
|\phi_{1}\rangle=|\!\uparrow\uparrow\rangle, 
\quad
|\phi_{2}\rangle=|\!\downarrow\downarrow\rangle,
\quad
|\phi_{3,4}\rangle=(|\!\uparrow\downarrow\rangle
\pm
|\!\downarrow\uparrow\rangle)/\sqrt{2}, 
$$
 with corresponding eigenvalues $\varepsilon_{\ell}$ given by $\varepsilon_{1}=\delta$, $\varepsilon_{2}=-\delta$, and $\varepsilon_{3,4}=0$. The Hamiltonian ${\cal H}_{E}(h(t)+\varepsilon_{\ell})$ in Eq.~(\ref{eq4}) is obtained from ${\cal H}_E$ by replacing $h(t)$ with $h_{\ell}(t)=h(t)+\varepsilon_{\ell}$ \cite{Yuan2007,Guo2016,Sun2007}. Consequently, the system’s evolution is governed by the dynamics of two effective Ising branches evolving under fields $h_{\ell}(t)=h(t)\pm\delta$, induced by the qubit-environment coupling.
%

%

%
The time-evolution operator for the total Hamiltonian ${\cal H}(t)$ can be written as  
\begin{equation}
\label{eq5}
U(t)=\sum_{\ell=1}^{4}|\phi_{\ell}\rangle\langle\phi_{\ell}|\otimes U_E(h_{\ell}(t)),   
\end{equation}
where $U_E(h_{\ell}(t))$ satisfies the time-dependent Schrödinger equation of the Ising chain in the presence of an effective magnetic field $h_{\ell}(t)$, namely  
\begin{equation}
\label{eq5a}
\frac{d}{dt}U_E(h_{\ell}(t))=-i  {\cal H}_E(h_{\ell}(t))\,U_E(h_{\ell}(t)).
\end{equation}
We assume that at the initial time $t=t_i$ the two central qubits $A$ and $B$ are uncorrelated with the environment. Hence the total density matrix is taken as $\rho(t_i)=\rho_{AB}(t_i)\otimes \rho_{E}(t_i)$. The initial state of the two central qubits is chosen to be a Werner state,  
\begin{equation}
\label{eq6}
\rho_{AB}(t_i)=\frac{1-a}{4}\,\mathds{1}+a|\phi_0\rangle\langle\phi_0|,   
\end{equation}
where $|\phi_0\rangle=(|\uparrow\uparrow\rangle+|\downarrow\downarrow\rangle)/\sqrt{2}$ is a Bell state, 
$\mathds{1}$ is the $4\times 4$ identity operator on the two–qubit Hilbert space, 
and $a \in [0,1]$ is the mixing parameter of the Werner state, interpolating between the maximally mixed state ($a=0$) and the pure Bell state $|\phi_0\rangle$ ($a=1$). 
The environment is initialized in its ground state $|\psi(t_i)\rangle_g$ at $t_i$.
Starting from this initial product state, the total system evolves unitarily under ${\cal H}(t)$. The density matrix at time $t$ is therefore  
\begin{equation}
\rho(t)=U(t)\,\rho(t_i)\,U^{\dagger}(t).
\end{equation}
The corresponding reduced density matrix of the two central qubits is obtained by tracing out the environment \cite{Liu2010,Yuan2007},  
%
%
\bea
\label{eq7}
\bl
\rho_{AB}(t)&={\rm Tr}_{E} [\rho(t)]
\\
&=
\sum_{\ell,\ell'=1}^{4} \langle\psi(t_i)|U^{\dagger}_{E}(h_{\ell'}(t))U_{E}(h_{\ell}(t))|\psi(t_i)\rangle \rho_{AB}(t_i)
\\
&=
\frac{1}{4} \left(
    \begin{array}{cccc}
      1+a & 0 & 0 & 2a\sqrt{|D(t)|} \\
      0 & 1-a & 0 & 0 \\
      0 & 0 & 1-a & 0 \\
      2a\sqrt{|D(t)|} & 0 & 0 & 1+a \\
    \end{array}
  \right),
  \no
  \el
  \\
\eea
%
where $D(t)= \langle\psi(h(t)+\delta)|\psi(h(t)-\delta)\rangle$ is the decoherence factor. Here, $|\psi(h(t)\pm\delta)\rangle=U_{E}(h(t)\pm\delta)|\psi(t_i)\rangle_{g}$ denote the environmental states at time $t$ evolved with the Hamiltonian ${\cal H}_E(h(t)\pm\delta)$.

Using the reduced density matrix in Eq.~(\ref{eq7}), the concurrence, which quantifies entanglement between the two qubits, is obtained as \cite{Liu2010}  
\begin{equation}
\label{eq8}
C=\max\Big[a\Big(\sqrt{|D(t)|}+\tfrac{1}{2}\Big)-\tfrac{1}{2},\,0\Big].
\end{equation}
In addition, the   quantum discord (QD) of the state $\rho_{AB}(t)$ can be expressed as \cite{Yuan2007}  
\begin{equation}
\label{eq9}
QD=\sum_{m=1}^{4}\lambda_{m}\log_{2}(\lambda_m)+f(a),
\end{equation}
where $\lambda_m$ are the eigenvalues of $\rho_{AB}(t)$, and  
%
$$
f(a) = 1 - \frac{1+a}{2}\log_{2}\!\left(\frac{1+a}{2}\right)
         - \frac{1-a}{2}\log_{2}\!\left(\frac{1-a}{2}\right).
$$
In the next sections, we analyze the dynamics of concurrence and quantum discord in both the absence and presence of the quantum reset process.

%
%
\begin{figure*}
\begin{minipage}{\linewidth}
\centerline{\includegraphics[width=0.33\linewidth]{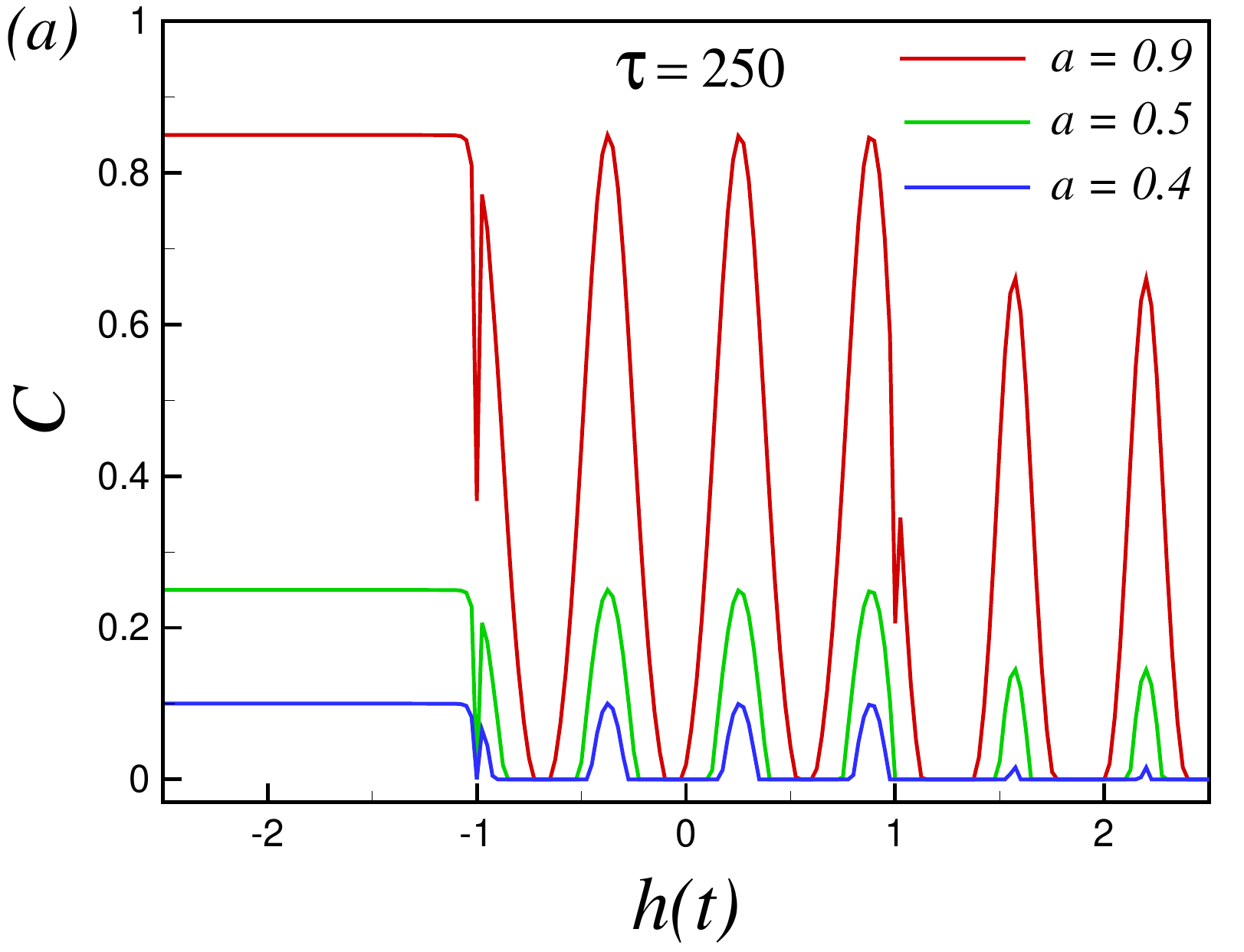}
\includegraphics[width=0.33\linewidth]{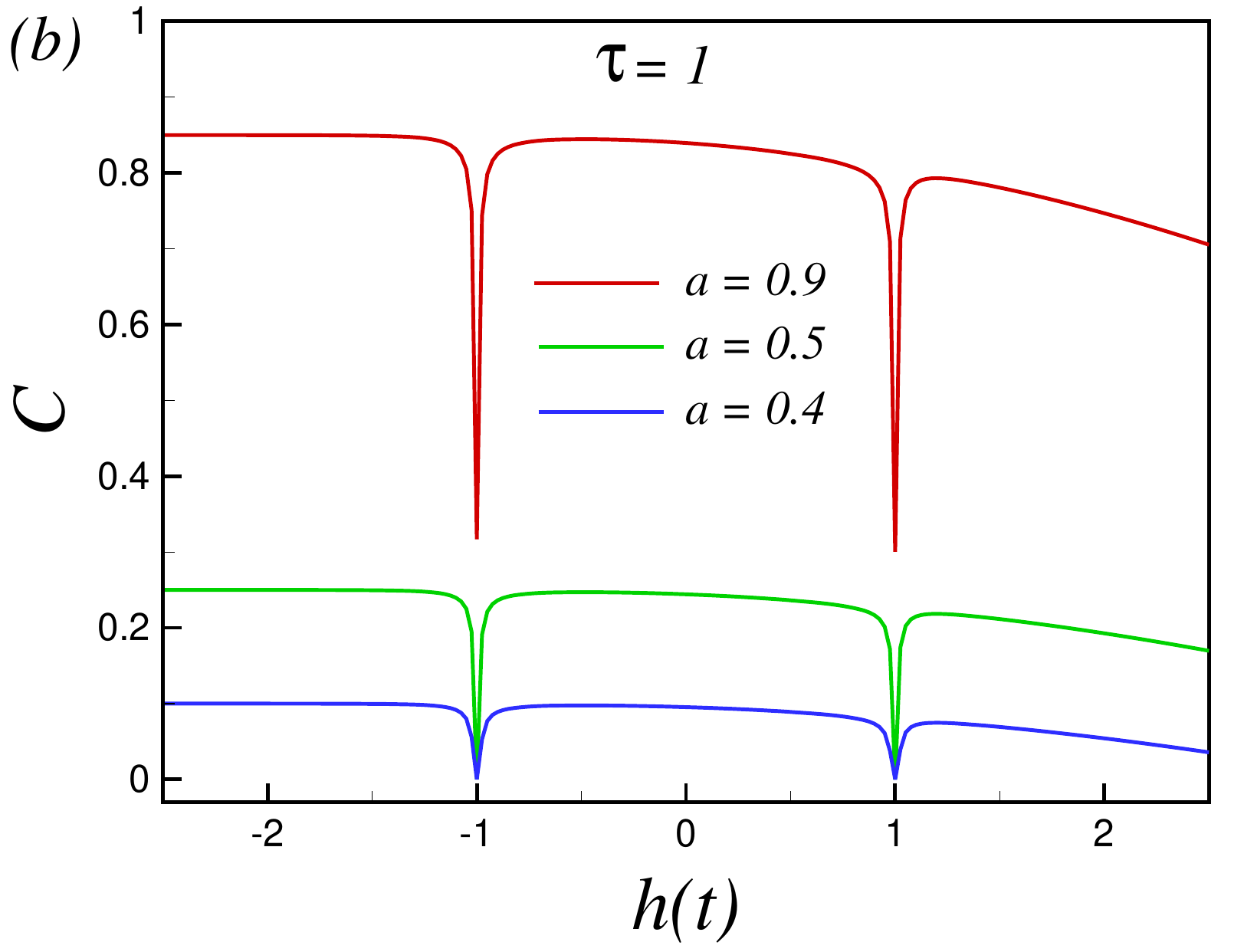}
\includegraphics[width=0.33\linewidth]{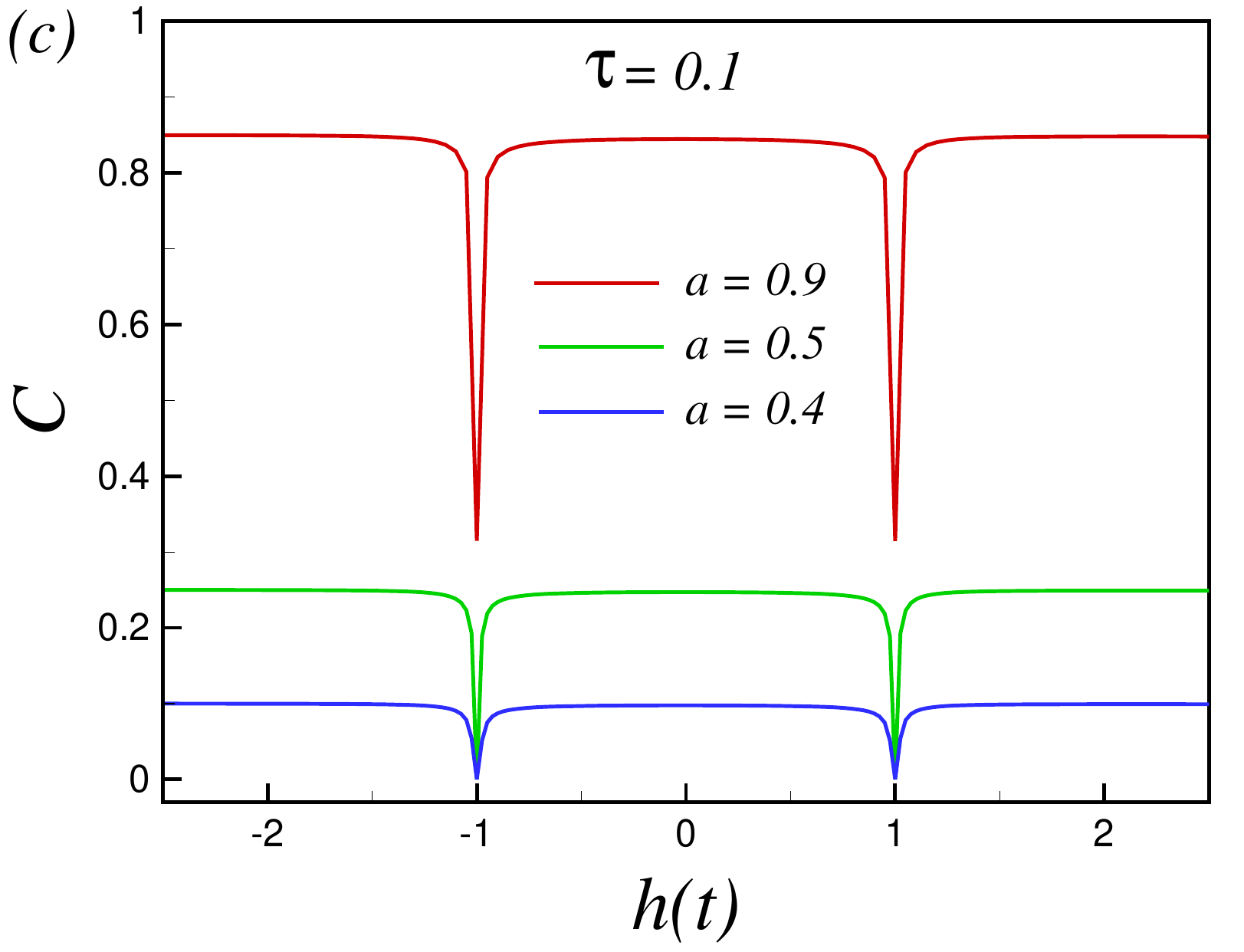}}
\centering
\end{minipage}
\caption{Concurrence ($C$) versus $h(t)=t/\tau$ in the absence of resetting, for chain size $N=500$, coupling $\delta=0.01$, and initial Werner parameter $a=0.9$.  
Panels correspond to different ramp times: (a) $\tau=250$ (strong coupling/slow ramp), (b) $\tau=1$, and (c) $\tau=0.1$ (weak coupling/fast ramp).  
In the slow-ramp regime, $C$ exhibits near-complete revivals between the critical points $h=\pm1$, whereas for shorter ramps it decays monotonically.
 }
\label{fig2}
\end{figure*}
%
%
\begin{figure*}
\begin{minipage}{\linewidth}
\centerline{\includegraphics[width=0.33\linewidth]{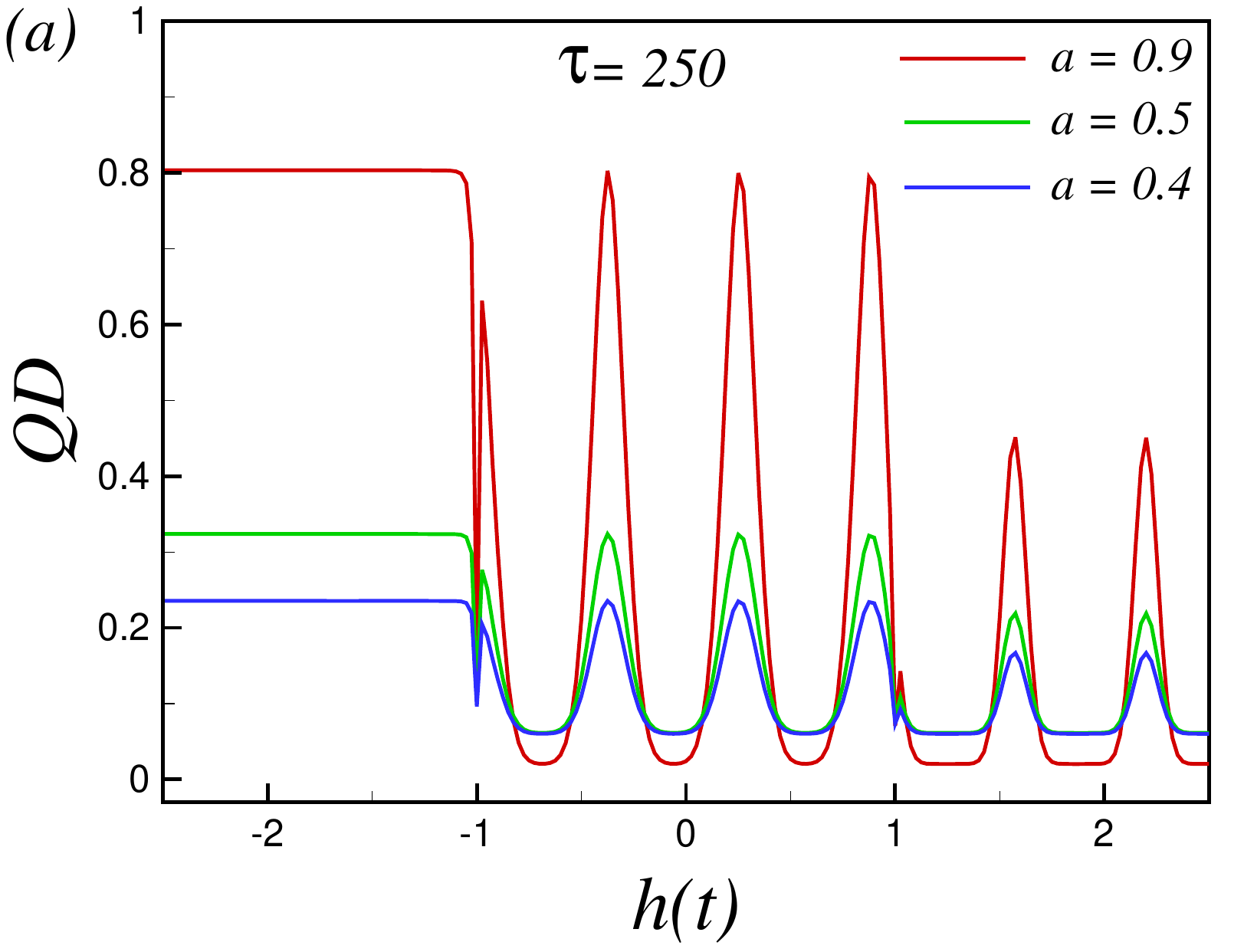}
\includegraphics[width=0.33\linewidth]{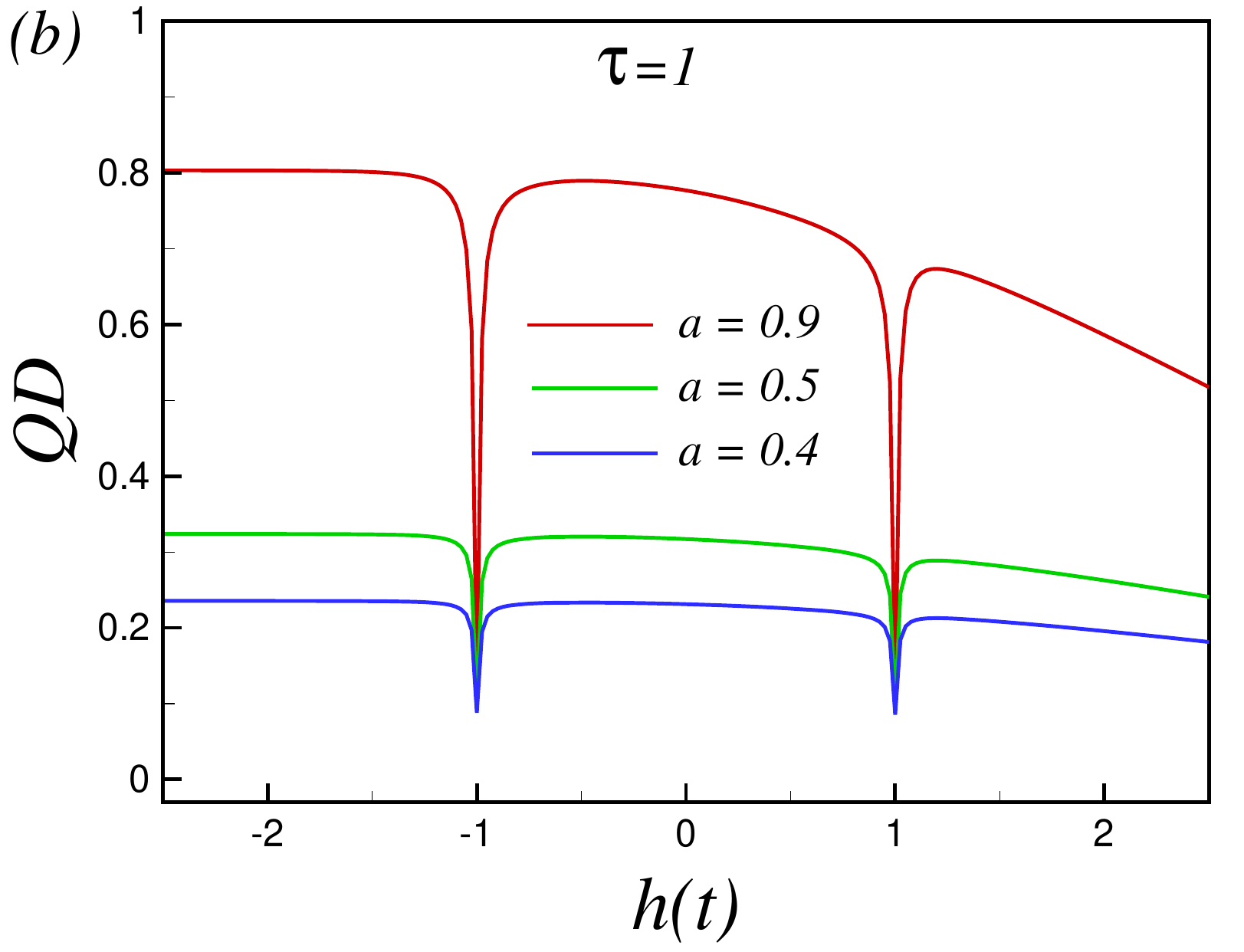}
\includegraphics[width=0.33\linewidth]{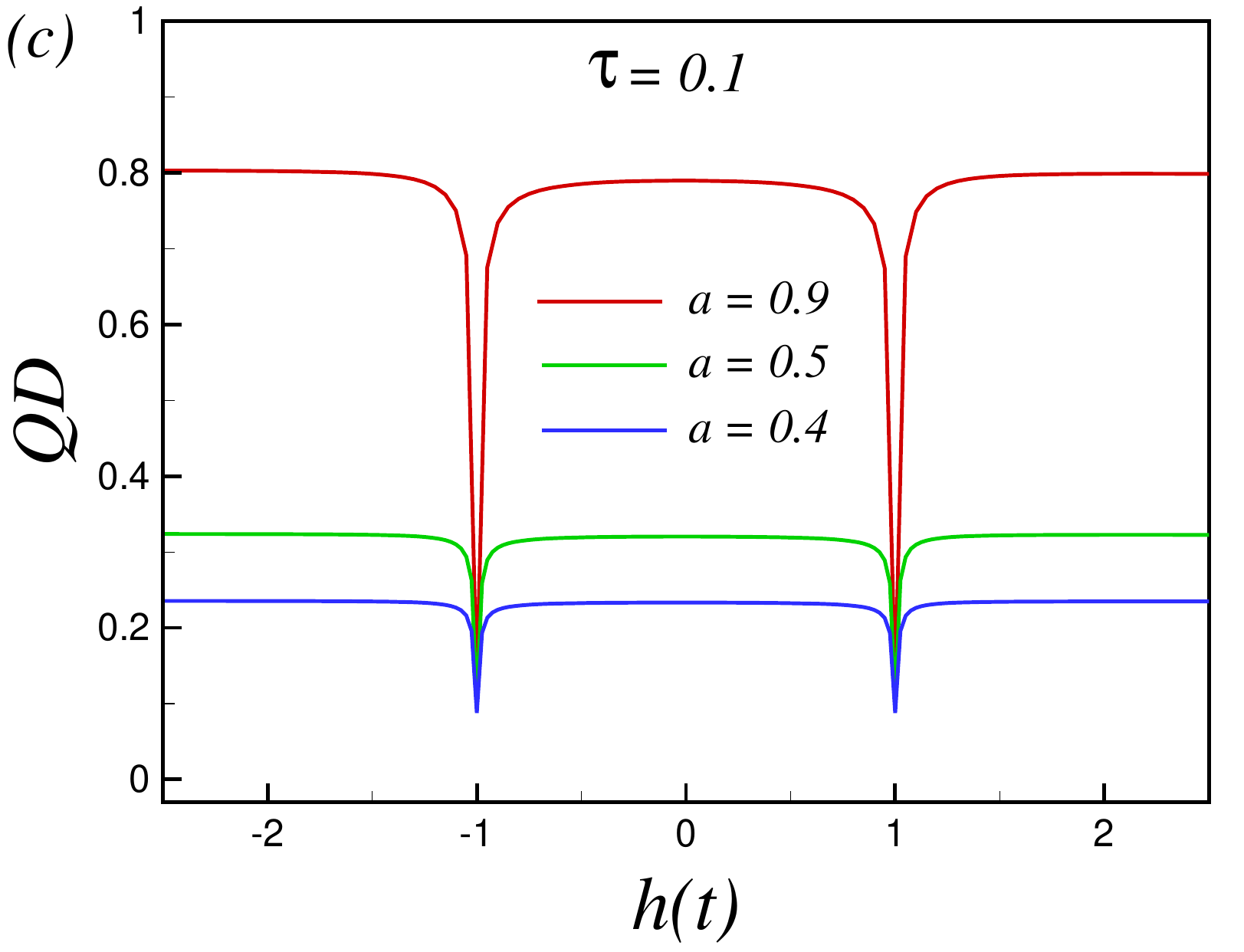}
}
\centering
\end{minipage}
\caption{Quantum discord (QD) versus $h(t)=t/\tau$ in the absence of resetting, with the same parameters as in Fig.~2.  
(a) For slow ramps ($\tau=250$), discord revives between the critical points, though less sharply than concurrence.  
(b,c) For intermediate and fast ramps ($\tau=1$, $\tau=0.1$), discord decays monotonically but remains finite even in regimes where concurrence vanishes.
}
\label{fig3}
\end{figure*}
%

\section{Reset-free dynamics}\label{sec:nqr}

In the absence of quantum reset, the decoherence factor can be obtained analytically 
by solving the time-dependent Schr\"{o}dinger equation,  Eq.~(\ref{eq5}).  
Applying the Jordan-Wigner transformation, followed by a Fourier transform \cite{Pfeuty1970,LSM1961,Jafari2012},  
the Hamiltonian of the environment [Eq.~(\ref{eq1})] with effective magnetic field $h(t)\pm\delta$ can be decomposed into a sum of $N/2$ noninteracting terms,  
\begin{equation}
{\cal H}_E(t) = \sum_{k} {\cal H}^{\pm}_{k}(t),
\end{equation}
where each mode Hamiltonian is given by  
\begin{equation}
\label{eq10}
\begin{aligned}
{\cal H}^{\pm}_{k}(t) &= \big(h(t)\pm\delta-\cos k\big)\,(c_{k}^{\dagger}c_{k}+c_{-k}^{\dagger}c_{-k}) \\
&\quad + \sin k \,(c_{k}^{\dagger}c_{-k}^{\dagger}+c_{k}c_{-k}),
\end{aligned}
\end{equation}
with wave numbers $k=(2m-1)\pi/N$ and $m=1,\dots,N/2$, where $N$ is the total number of spins in the chain.

Equation~(\ref{eq10}) shows that the Hamiltonian of $N$ interacting spins [Eq.~(\ref{eq1})] can be mapped to a set of $N/2$ noninteracting quasispins.  
The Bloch single-particle Hamiltonian ${\cal H}^{\pm}_{k}(t)$ takes the form  
\begin{equation}
\label{eq11}
{\cal H}^{\pm}_{k}(t)=
\begin{pmatrix}
h_{z}(k,t) & h_{x}(k) \\
h_{x}(k)   & -h_{z}(k,t)
\end{pmatrix},
\end{equation}
with $h_{x}(k)=\sin k$ and $h_{z}(k,t)=h(t)\pm\delta - \cos k$.  
Defining the state vector $|\psi^{\pm}_{k}(t)\rangle = |\psi_{k}(h(t)\pm\delta)\rangle = \big(v^{\pm}_{k}(t),\, u^{\pm}_{k}(t)\big)^{T}$,  
the time-dependent Schr\"{o}dinger equations become  
\begin{eqnarray}
\label{eq12}
\bl
i\frac{d}{dt}v^{\pm}_{k} &= -\big(h(t)\pm\delta-\cos k\big)\, v^{\pm}_{k} + \sin k\, u^{\pm}_{k}, \\
i\frac{d}{dt}u^{\pm}_{k} &= \phantom{-}\big(h(t)\pm\delta-\cos k\big)\, u^{\pm}_{k} + \sin k\, v^{\pm}_{k}. 
\el
\end{eqnarray}
These equations are exactly solvable \cite{Damski2005,Damski2011,Suzuki2016,Jafari2024a,Nag2012,Ansari:2026,Kheiri:2025aa} (see Appendix~\ref{APA}).  
Within this formalism, the decoherence factor is given by  
\begin{equation}
\label{eq9}
D(t)= \prod_{k>0} F_k(t); 
\quad 
F_k(t) = \big|\,u_{k}^{+*}(t)u_{k}^-(t) + v_{k}^{+*}(t)v_{k}^-(t)\,\big|^2.
\end{equation}
The amplitudes $u^{\pm}_{k}(t)$ and $v^{\pm}_{k}(t)$ can also be obtained numerically from the von Neumann equation,  
\begin{equation}
\dot{\rho}^{\pm}_{k}(t)=-i\,[{\cal H}^{\pm}_{k}(t),\rho^{\pm}_{k}(t)],
\end{equation}
where $\rho^{\pm}_{k}(t)$ is the density operator for mode $k$. In the instantaneous eigenbasis of ${\cal H}^{\pm}_{k}(t)$, the matrix elements are  
\begin{eqnarray} 
\label{eq:amplitudes}
\bl
\rho^{(d)\pm}_{k,11}(t) &= |v^{\pm}_{k}(t)|^2 = {_g}\langle \psi^{\pm}_{k}(t)|\rho^{\pm}_{k}(t)|\psi^{\pm}_{k}(t)\rangle_g, 
\\
\rho^{(d)\pm}_{k,12}(t) &= v^{\pm}_{k}(t)u^{\ast\pm}_{k}(t) = {_g}\langle \psi^{\pm}_{k}(t)|\rho^{\pm}_{k}(t)|\psi^{\pm}_{k}(t)\rangle_{e},  
\\
\rho^{(d)\pm}_{k,21}(t) &= v^{\ast\pm}_{k}(t)u^{\pm}_{k}(t) = {_e}\langle \psi^{\pm}_{k}(t)|\rho^{\pm}_{k}(t)|\psi^{\pm}_{k}(t)\rangle_{g}, 
\quad\quad
 \\
\rho^{(d)\pm}_{k,22}(t) &= |u^{\pm}_{k}(t)|^2 = {_e}\langle \psi^{\pm}_{k}(t)|\rho^{\pm}_{k}(t)|\psi^{\pm}_{k}(t)\rangle_{e}, 
\el
\end{eqnarray}
where $|\psi^{\pm}_{k}(t)\rangle_g$ and $|\psi^{\pm}_{k}(t)\rangle_e$ denote the instantaneous ground and excited states of ${\cal H}^{\pm}_{k}(t)$, respectively.  
In this representation, Eq.~(\ref{eq9}) can be expressed in density-matrix form as  
\begin{equation}
\label{eq16}
F_k(t) 
\!=
\rho^{(d)+}_{k,11}\rho^{(d)-}_{k,11}
+\rho^{(d)+}_{k,22}\rho^{(d)-}_{k,22}
+\rho^{(d)+}_{k,12}\rho^{(d)-}_{k,21}
+\rho^{(d)+}_{k,21}\rho^{(d)-}_{k,12}.
\end{equation}
This formulation is valid for both pure and mixed states. In particular, it remains applicable when the system evolves under quantum reset, where the density matrix is generally mixed.


\section{Dynamics with Quantum Reset}\label{sec:qr}

We now consider the reset dynamics of the environment, initially prepared at $t_i$ in its ground state $|\psi(t_i)\rangle_{g}$. Its evolution over an infinitesimal interval $dt$ is  
\begin{equation}
\bl
\no
\label{eq:resetSchrodinger}
|\psi(t+dt)\rangle =
\begin{cases} 
|\psi(t_i)\rangle_g, & {\rm with\  prob.=}\; r dt, \\[6pt]
\big[1 - i\,{\cal H}_E(t)\,dt\big]\,|\psi(t_i)\rangle_g, & {\rm with\ prob.=}\; 1-r dt,
\end{cases}
\el
\\
\end{equation}
where $r \ge 0$ is the reset rate. Thus, within a short interval $dt$, the system either reverts to its initial state with probability $r\,dt$ or evolves unitarily under ${\cal H}_E(t)$ with complementary probability $1-r\,dt$.  
We assume that the resetting is a Poisson process with rate $r$, therefore the probability that there is no reset is simply $e^{-rt}$
and the probability of a reset is just $rdt$. Hence, taking the product, the probability of this event is $re^{-rt}dt$.
For $r=0$, the evolution is purely unitary. For $r>0$, the dynamics combine stochastic and deterministic contributions, and the density matrix becomes history-dependent, reflecting fluctuations across different reset realizations. The physically observed density matrix is obtained by averaging over all possible reset histories.

The reset-free density matrix at time $t$ is  
\begin{equation}
\rho_{0}(t)=|\psi(t)\rangle\langle\psi(t)|=\prod_{k}\rho_{k,0}(t),
\end{equation}
assuming a pure state. The ensemble-averaged density matrix under reset is then given by \cite{Evans2014,Majumdar2015,Mukherjee2018,Magoni2020,Sevilla2023}  
\begin{eqnarray}
\label{eq:QRdensitymatrix}
\rho_{r}(t) &=& \prod_{k}\rho_{k,r}(t), 
\end{eqnarray}
with
\begin{eqnarray}
\rho_{k,r}(t) &=& r \int_{0}^{t} e^{-r t'}\, \rho_{k,0}(t')\, dt' + e^{-rt}\,\rho_{k,0}(t), 
\end{eqnarray}
where the integral term accounts for all trajectories with at least one reset before $t$, and the exponential term corresponds to the case with no reset within $[0,t)$.

%
\begin{figure*}
\begin{minipage}{\linewidth}
\centerline{\includegraphics[width=0.33\linewidth]{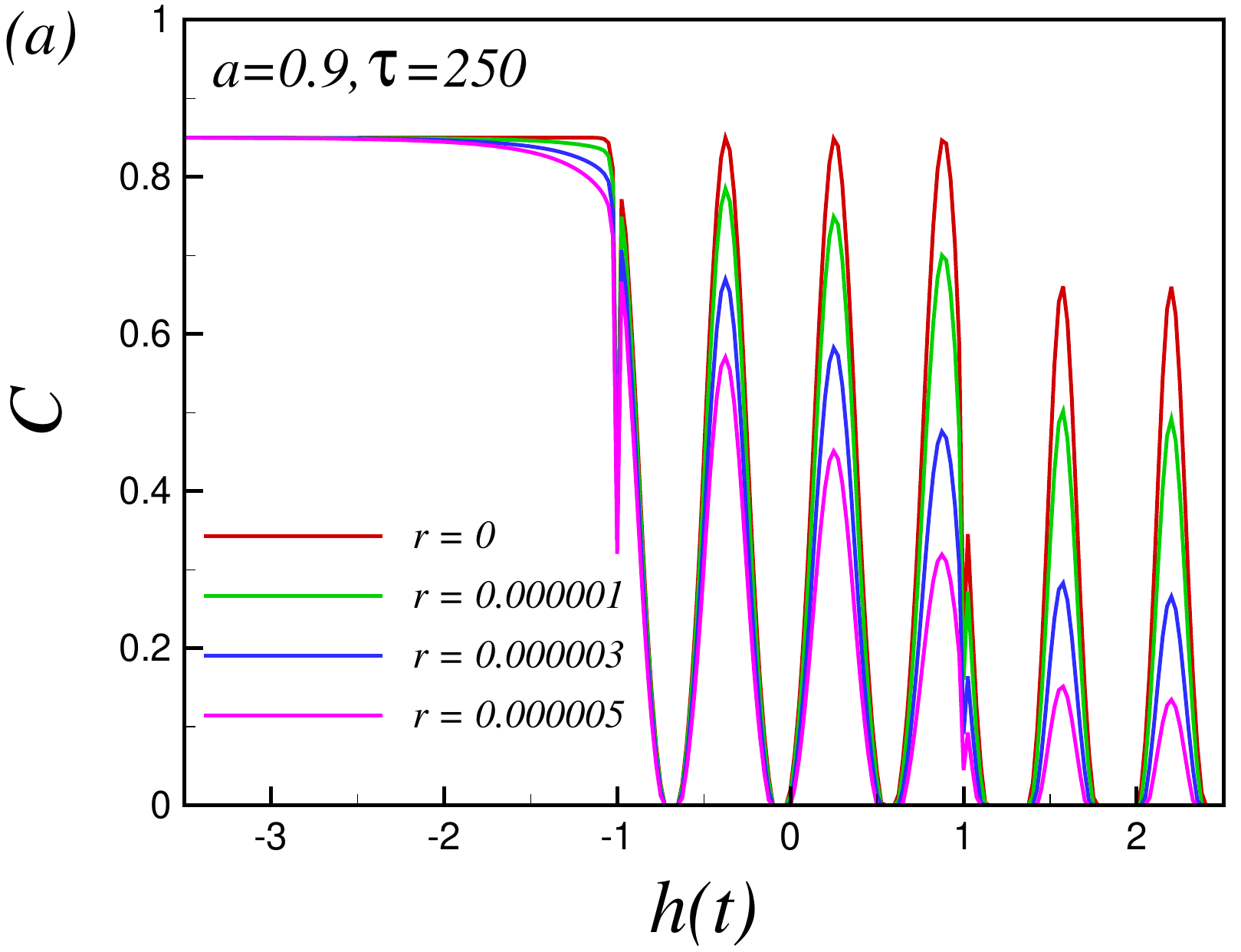}
\includegraphics[width=0.33\linewidth]{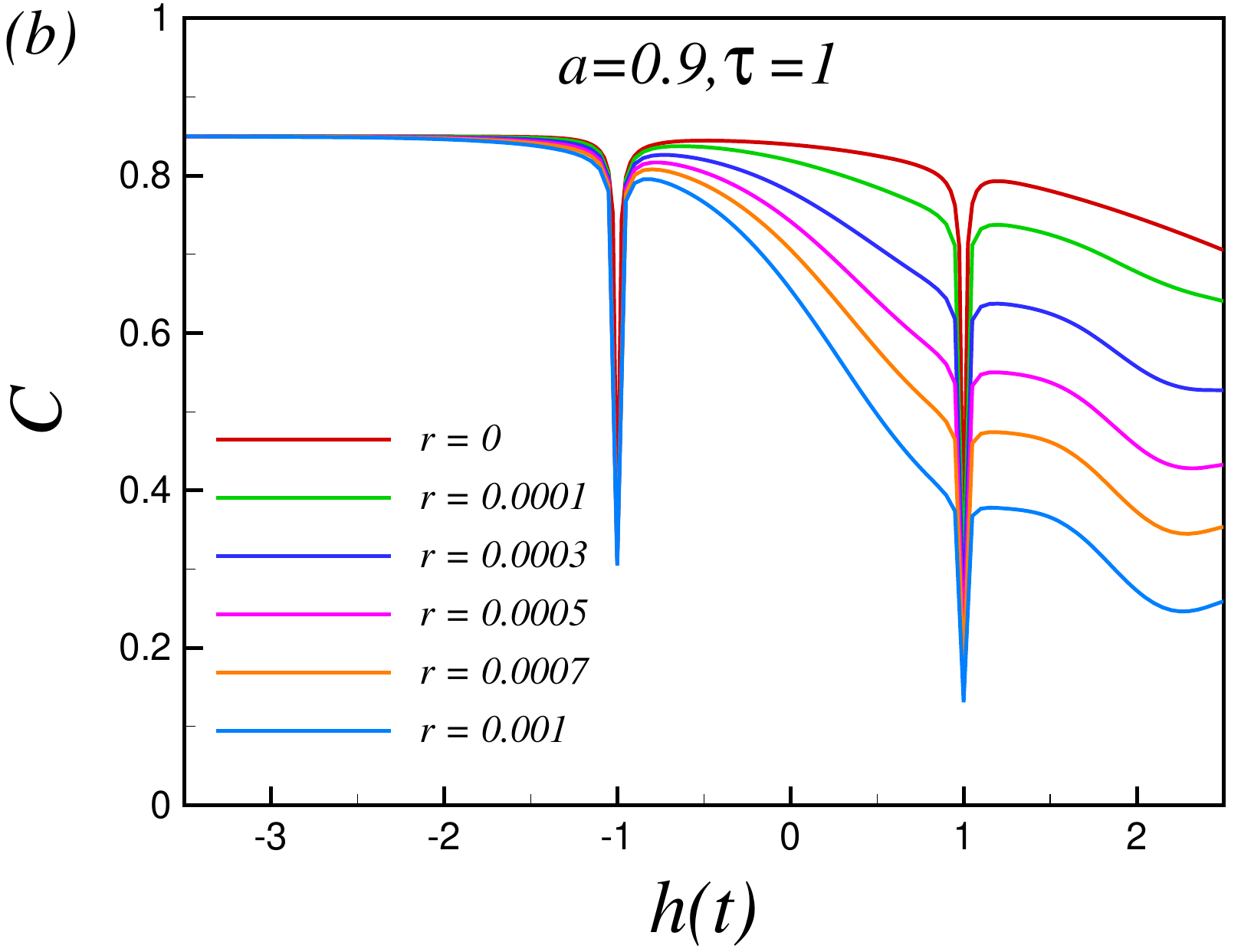}
\includegraphics[width=0.33\linewidth]{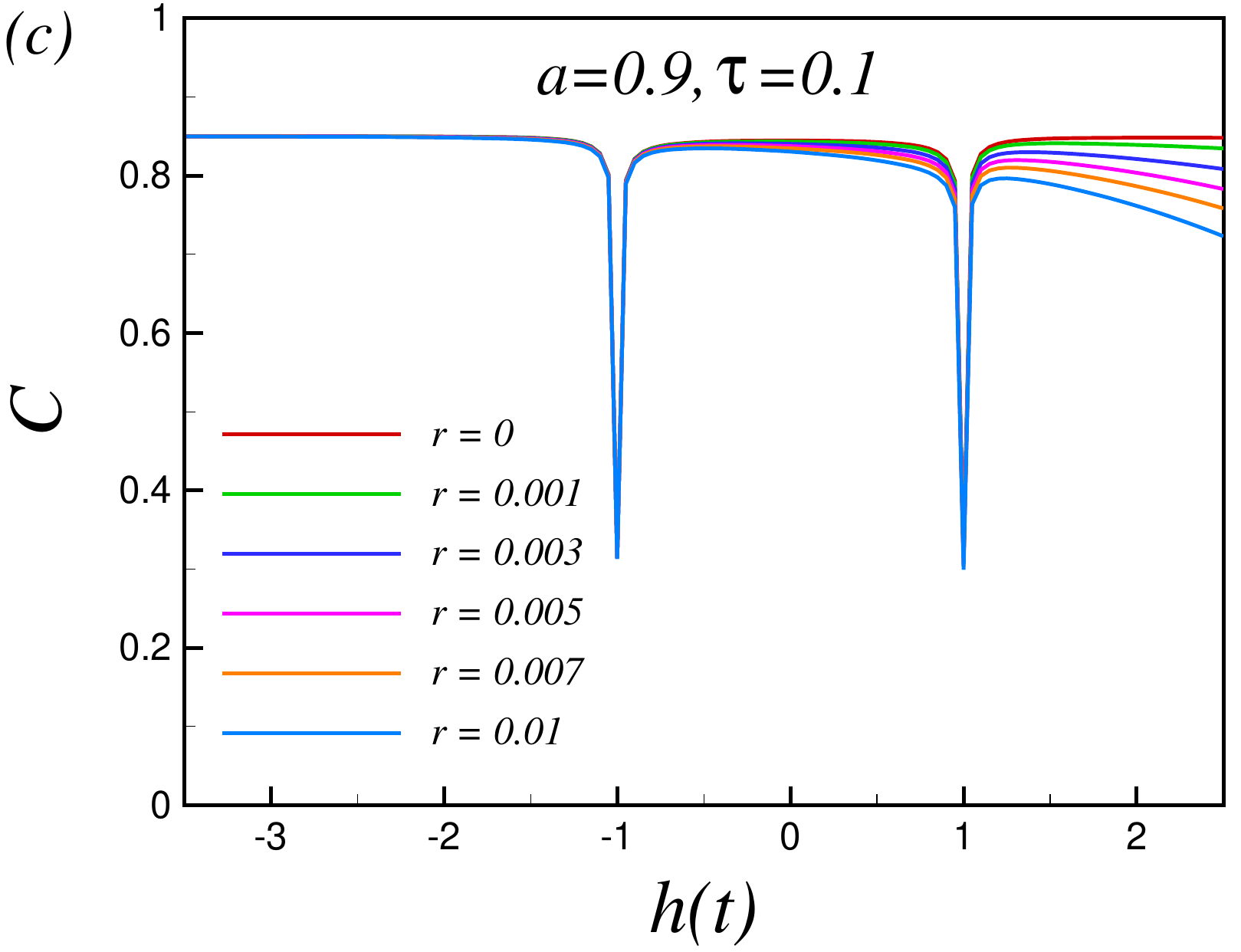}
}
\centering
\end{minipage}
\caption{Concurrence versus $h(t)=t/\tau$ under stochastic resetting of the environment, for $N=500$, $\delta=0.01$, $a=0.9$, and reset rates $r$ as indicated.  
(a) Slow ramp ($\tau=250$): partial revivals persist between $h=\pm1$ but their amplitudes decrease with increasing $r$.  
(b) Intermediate ramp ($\tau=1$): revivals are suppressed and correlations decay more rapidly with $r$.  
(c) Fast ramp ($\tau=0.1$): correlations are strongly reduced and only weakly dependent on $r$.
}
\label{fig4}
\end{figure*}
%
%
\begin{figure*}
\begin{minipage}{\linewidth}
\centerline{\includegraphics[width=0.33\linewidth]{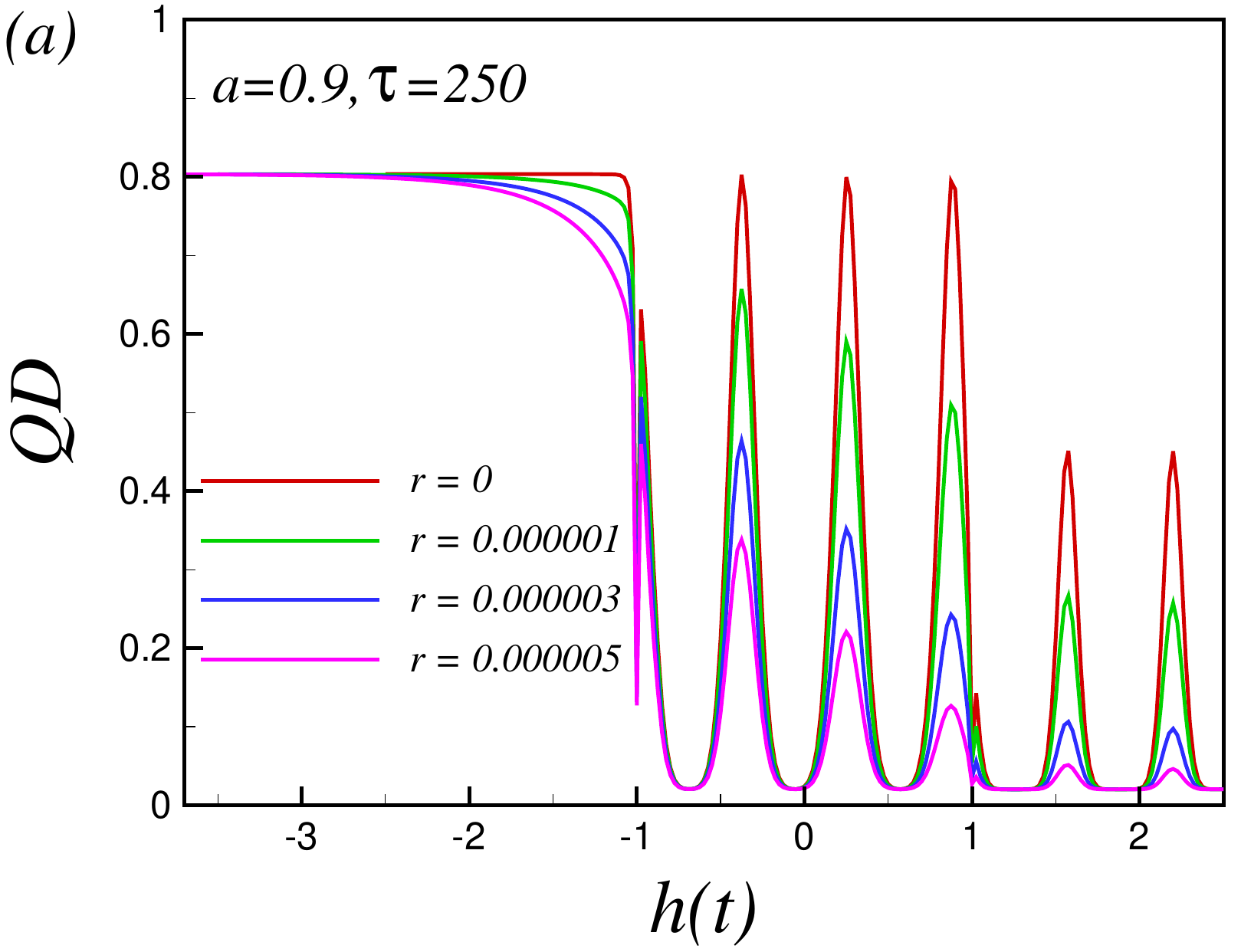}
\includegraphics[width=0.33\linewidth]{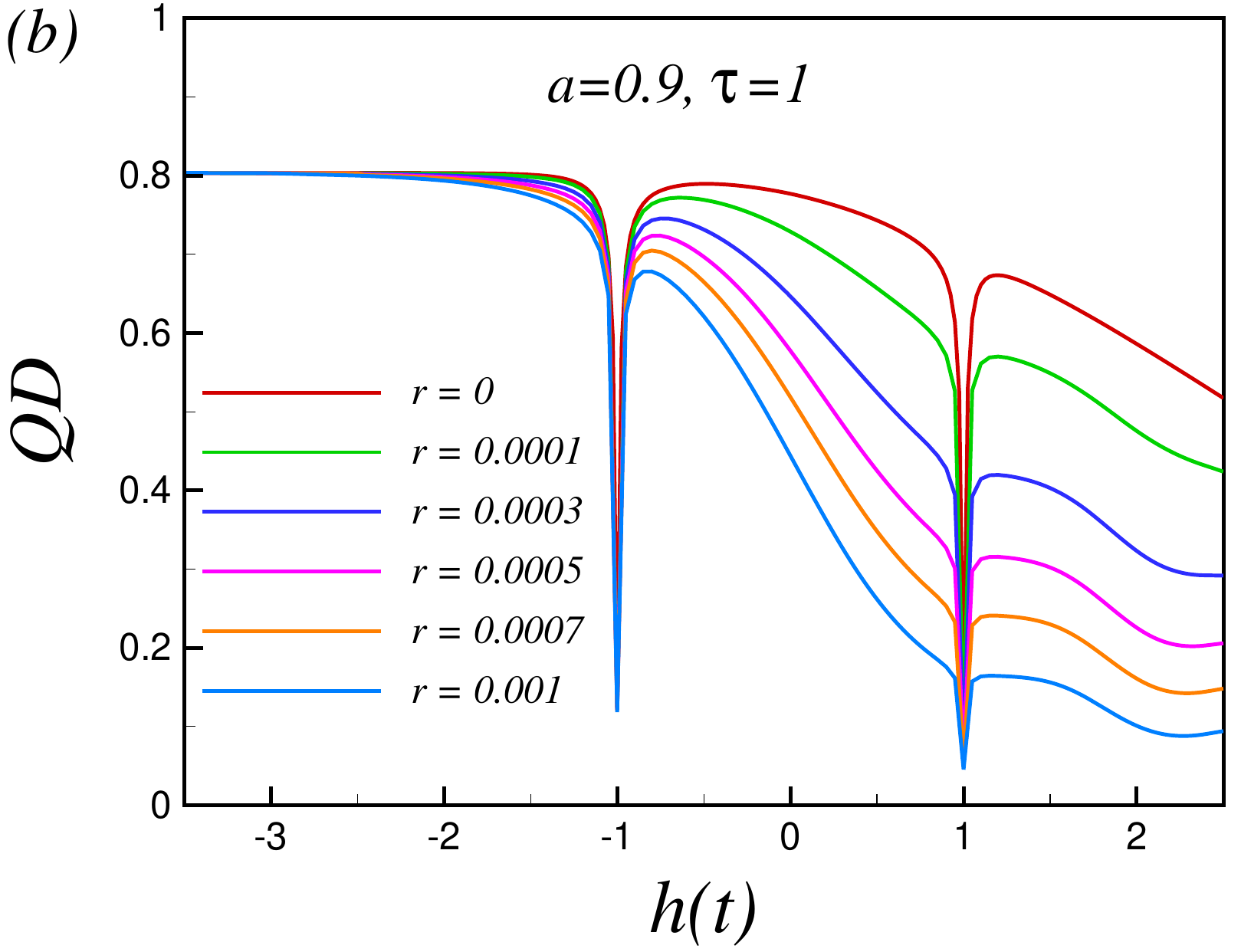}
\includegraphics[width=0.33\linewidth]{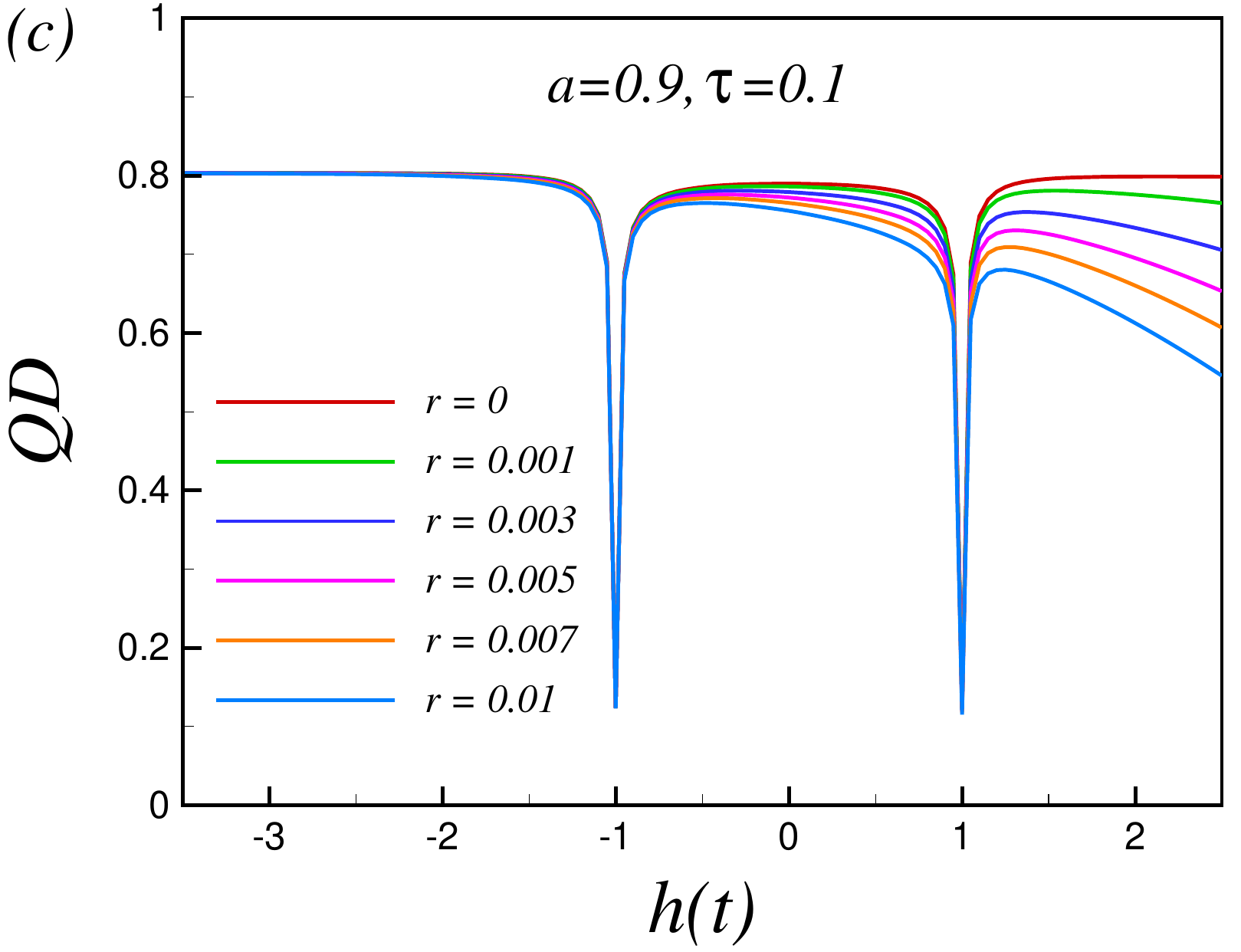}
}
\centering
\end{minipage}
\caption{Quantum discord  versus $h(t)=t/\tau$ under stochastic resetting, for the same parameters as in Fig.~4.  
Resetting suppresses revivals and accelerates decay, although discord remains finite even when concurrence vanishes.  
The effect of increasing $r$ is most visible in the slow-ramp case (a), while for faster ramps (b,c) the dynamics are dominated by monotonic suppression.
}
\label{fig5}
\end{figure*}
%

\section{Numerical Analysis of Correlation Dynamics}\label{sec:num}

In this section we present the results of our numerical simulations, carried out using an analytical framework, to analyze the dynamics of entanglement and quantum discord for two qubits coupled to a linearly driven transverse-field Ising chain. We compare the cases with and without the quantum reset process.  
We study the behavior of concurrence and quantum discord for different ramp time scales $\tau$, as functions of the initial-state parameter $a$ and the instantaneous transverse field $h(t)$, while the environment is driven across its quantum critical points.  
The system is initially prepared in the ground state of the Ising chain at $h_i=-5$. The transverse field is then ramped up so that the system evolves through the ferromagnetic phase ($-1 < h(t) < 1$) and subsequently enters the paramagnetic phase ($h(t) > 1$). In this process, the quench necessarily crosses the two critical points located at $h_c=\pm 1$.

\subsection{Correlation Dynamics without Quantum Reset}

In the absence of resetting, the decoherence factor in the paramagnetic phase can be approximated as~\cite{Damski2011,Suzuki2016,Nag2012}  
\begin{equation}
D(t) \approx \exp\!\left[-\frac{N\delta^2}{4h(t)^2\big(h(t)^2-1\big)}\right].
\label{eq13}
\end{equation}
The corresponding profiles of concurrence $C$ and quantum discord QD are shown in Fig.~\ref{fig2} and Fig.~\ref{fig3}, as functions of $h(t)=t/\tau$ for ramp time scales $\tau=250$, $\tau=1$, and $\tau=0.1$.  

Our analysis reveals that when the initial-state parameter lies in the interval $a \in [0,1/3]$, the two central qubits remain unentangled and the concurrence vanishes identically, although the discord can still take nonzero values. To illustrate the dynamical behavior more clearly, we therefore focus on three representative cases: $a=0.4$, $a=0.5$, and $a=0.9$.
As expected, the adiabatic evolution breaks down once the transverse field crosses the first critical point at $h_c=-1$, where the energy gap closes. This breakdown leads to accelerated decoherence, which manifests as a reduction in quantum correlations~\cite{Damski2011,Suzuki2016,Nag2012}.  
The dynamics of decoherence are mainly influenced by two factors: (i) the enhanced sensitivity of the environment near the quantum critical points, which amplifies the effect of perturbations~\cite{Zamani2024,Jafari2024a}, and (ii) the excitations generated in the environment as the system is driven through the critical point.

As seen from Fig.~\ref{fig2} and Fig.~\ref{fig3}, both concurrence and QD remain essentially unchanged until the transverse field reaches the first critical point at $h_c=-1$, reflecting adiabatic evolution. At this point, the band gap closes for $k=\pi$, and the breakdown of adiabaticity is signaled by an abrupt change in both measures. Beyond $h_c=-1$, the correlations either undergo nearly perfect revivals [Figs.~\ref{fig2}(a),~\ref{fig3}(a)] or decay monotonically [Figs.~\ref{fig2}(b)-(c),~\ref{fig3}(b)-(c)] within the region bounded by the two Ising critical points. These contrasting behaviors originate from non-adiabatic dynamics near criticality, where large-$k$ modes are strongly excited. In particular, modes with $k \sim \pi - \hat{k}$, with $\hat{k}\sim\tau^{-1/2}$, exhibit significant excitation~\cite{Damski2011,Suzuki2016,Nag2012}, while small-$k$ modes with $k \ll \pi - \hat{k}$ evolve adiabatically across the transition~\cite{Damski2011,Suzuki2016,Nag2012}.

Complete revivals of concurrence and QD between the two critical points are observed when the qubit–environment coupling is sufficiently strong, i.e., $\delta \gg \pi/(16\tau)$, as illustrated in Figs.~\ref{fig2}(a) and \ref{fig3}(a) for $\tau=250$~\cite{Damski2011,Suzuki2016,Nag2012}. These revivals occur in the magnetic-field domain $h(t)$ with a characteristic period $\pi/(4\tau\delta)$, identical to that of the decoherence factor~\cite{,Suzuki2016,Nag2012}. In contrast, in the weak-coupling regime $\delta \ll \pi/(16\tau)$, both concurrence and QD decay monotonically, as shown in Figs.~\ref{fig2}(b)-(c) and Figs.~\ref{fig3}(b)-(c) for $\tau=1$ and $\tau=0.1$.

Moreover, as the driven field reverses polarity and crosses the second critical point at $h_c=1$, both concurrence and QD exhibit a monotonic and accelerated reduction. At this point, the energy gap closes at $k=0$, and modes with $k \sim \hat{k}$ become significantly excited~\cite{Damski2011,Suzuki2016,Nag2012}. The breakdown of adiabatic evolution at $h_c=1$ is thus reflected in sharp changes of both concurrence and QD. It is worth noting, however, that unlike concurrence, quantum discord between the two central qubits never vanishes, even when the initial state parameter satisfies $a<1/3$, corresponding to a completely mixed initial state of the qubits.

%
\begin{figure}
\begin{minipage}{\columnwidth}
\centerline{\includegraphics[width=0.5\columnwidth]{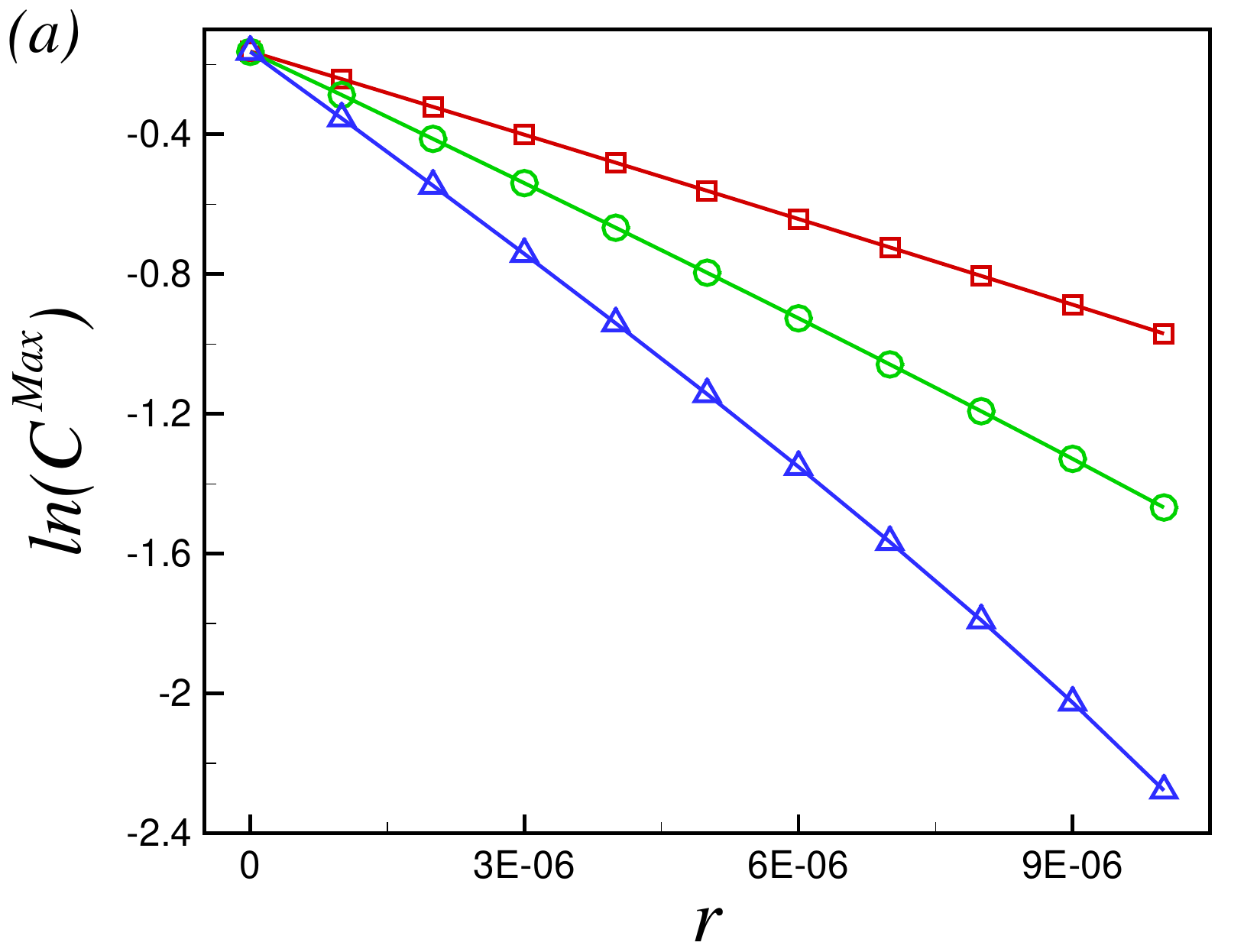}
\includegraphics[width=0.5\columnwidth]{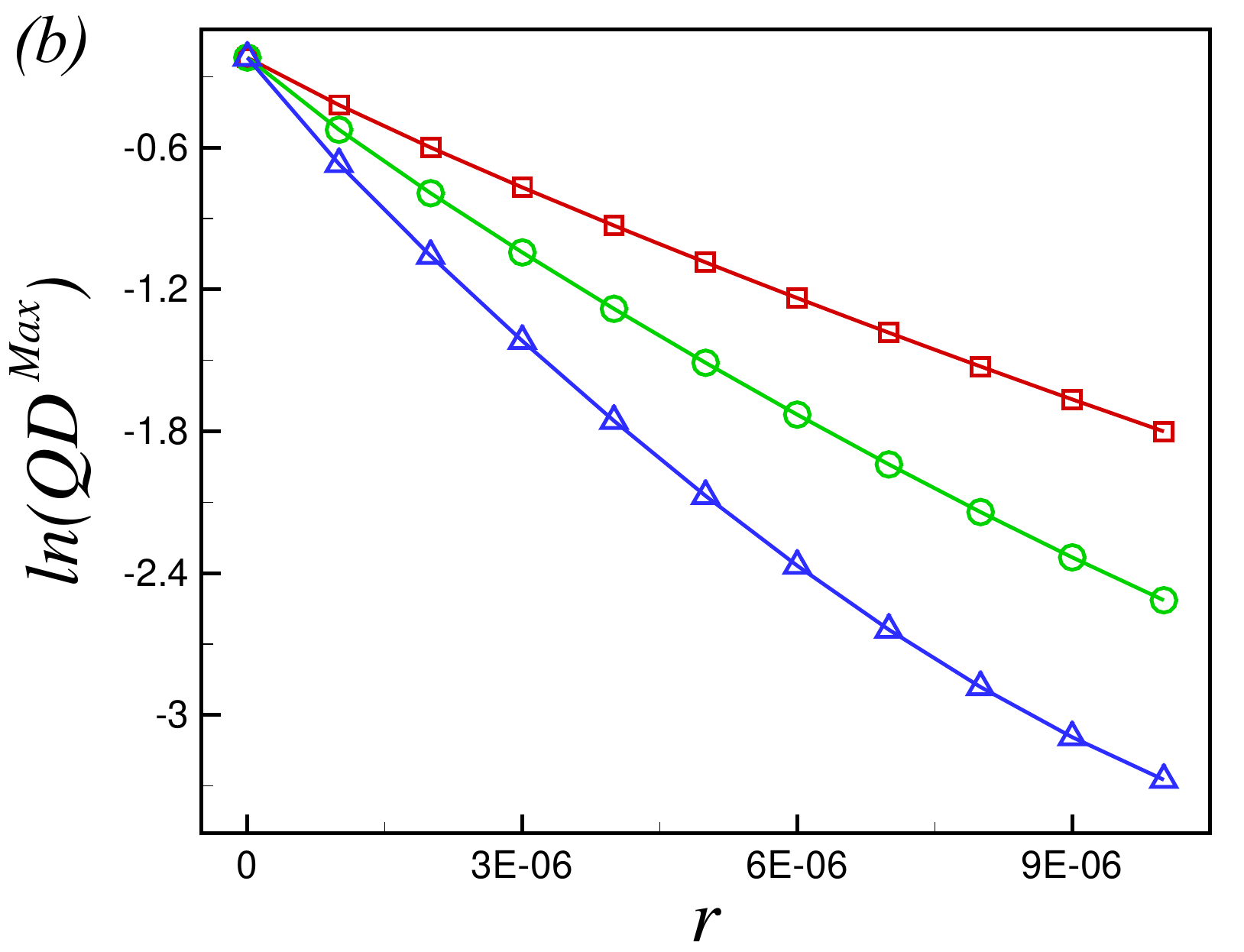}}
\centering
\end{minipage}
\caption{Scaling of peak correlations with reset rate $r$, for $N=500$, $\delta=0.01$, $a=0.9$, and $\tau=250$.  
(a) Peak concurrence C$^{\rm Max}$ decays exponentially with $r$, as seen from the linear dependence of  ln[C$^{\rm Max}$] on $r$.  
(b) Peak discord QD$^{\rm Max}$ shows no clear scaling with $r$, highlighting the different sensitivity of the two measures.
}
\label{fig6}
\end{figure}
%

\subsection{Dynamics under Quantum Reset}
We now examine the effect of quantum resetting of the environment on the dynamics of quantum correlations.  
For clarity we focus on the representative case $a=0.9$ (without loss of generality), noting that for $a<1/3$ the entanglement remains zero even in the presence of resetting.  
The influence of resetting is illustrated in Figs.~\ref{fig4} and \ref{fig5} for ramp time scales (a)~$\tau=250$, (b)~$\tau=1$, and (c)~$\tau=0.1$.  
Unlike the reset-free case, concurrence and discord decrease even before the transverse field reaches the first critical point at $h(t)=-1$, reflecting the non-adiabatic character of the evolution induced by resetting.  

In the strong-coupling regime ($\tau=250$), shown in Figs.~\ref{fig4}(a) and \ref{fig5}(a), partial revivals of concurrence and discord persist under resetting, but their amplitudes decrease with increasing reset rate $r$.  
This behavior contrasts with the reset-free case, where revivals between the critical points remain robust as the ramp time increases.  
The decay of revivals in the presence of resetting originates from additional excitations introduced during the evolution~\cite{Jafari2025g}.  
Notably, the revival period remains unchanged in both scenarios.  
Numerical analysis further reveals that the peak concurrence decays exponentially with $r$ [Fig.~\ref{fig6}(a)], whereas the peak discord shows no clear scaling [Fig.~\ref{fig6}(b)].

The exponential sensitivity of concurrence to resetting can be traced back to the
structure of the decoherence factor, which is expressed as a product of overlap
terms across all momentum modes [Eq.~(\ref{eq9})]. Since resetting interrupts coherent
mode evolution, the cumulative effect is to introduce an exponential suppression
of revival peaks. In contrast, quantum discord includes both classical and
non-classical contributions to total correlations, which makes it less sensitive
to phase-coherent interference effects. This difference explains why no clear
scaling emerges for QD, in agreement with our numerical findings.

\begin{figure}[t!]
\centering
\includegraphics[width=0.495\linewidth]{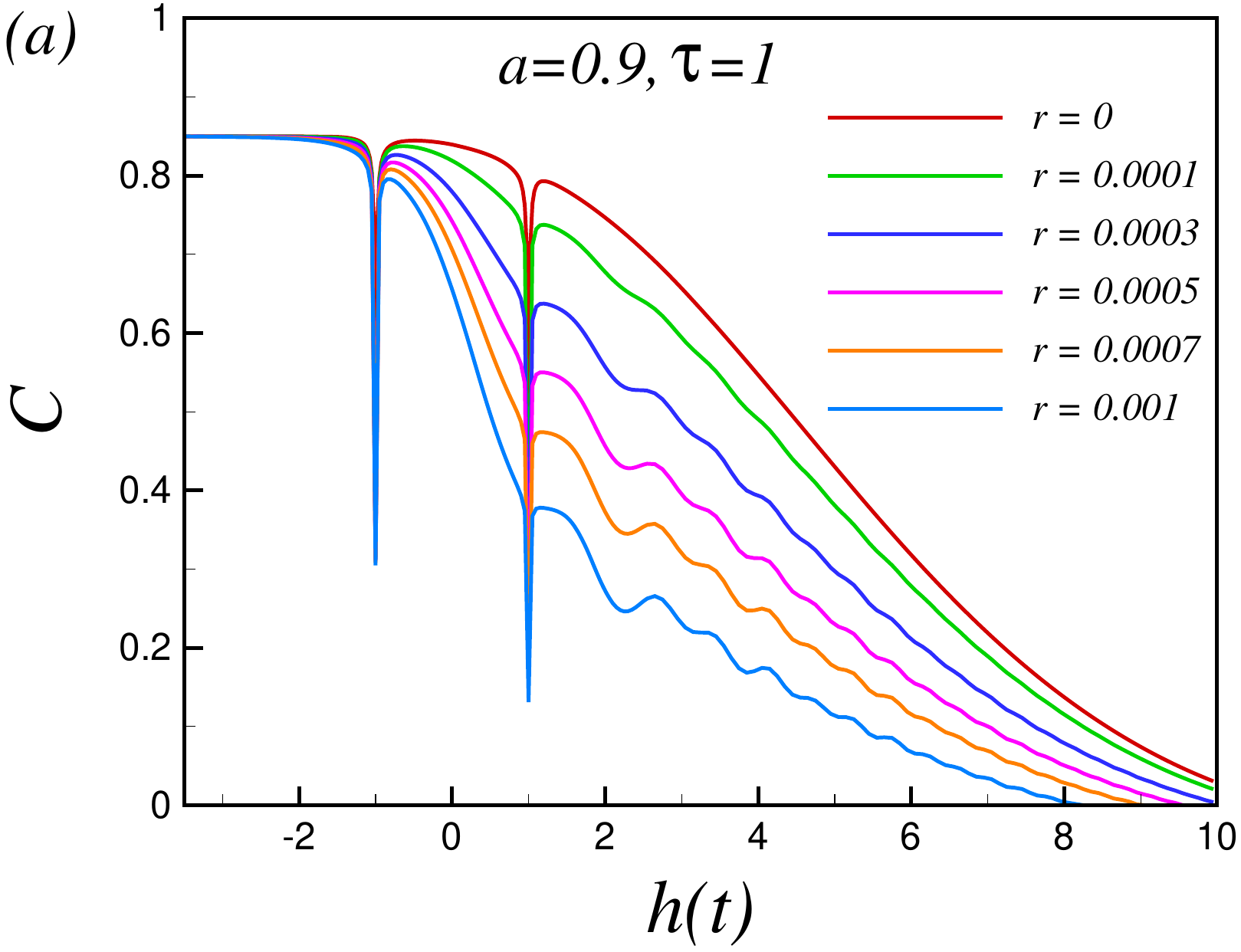}
\includegraphics[width=0.495\linewidth]{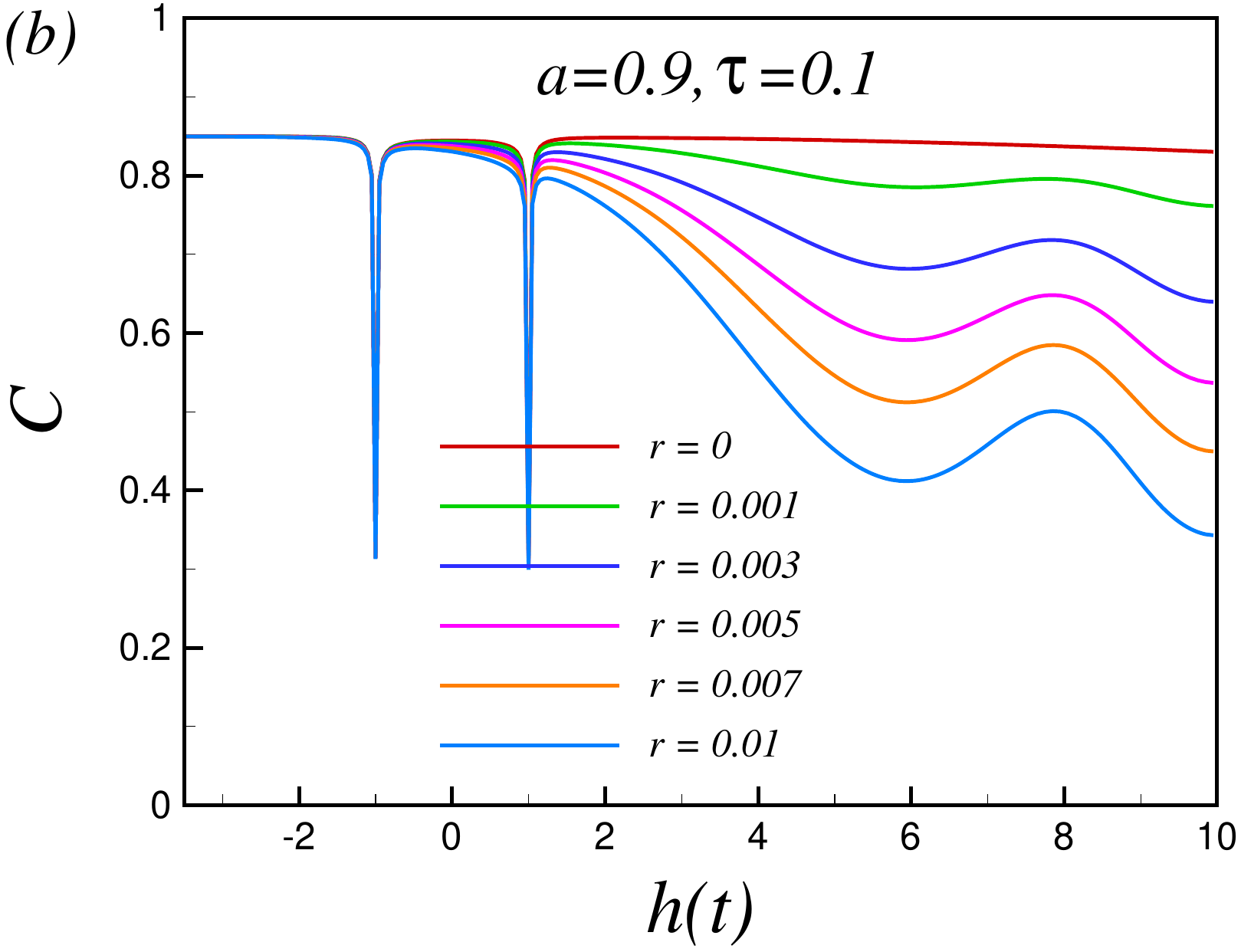}
\\[4pt]
\includegraphics[width=0.495\linewidth]{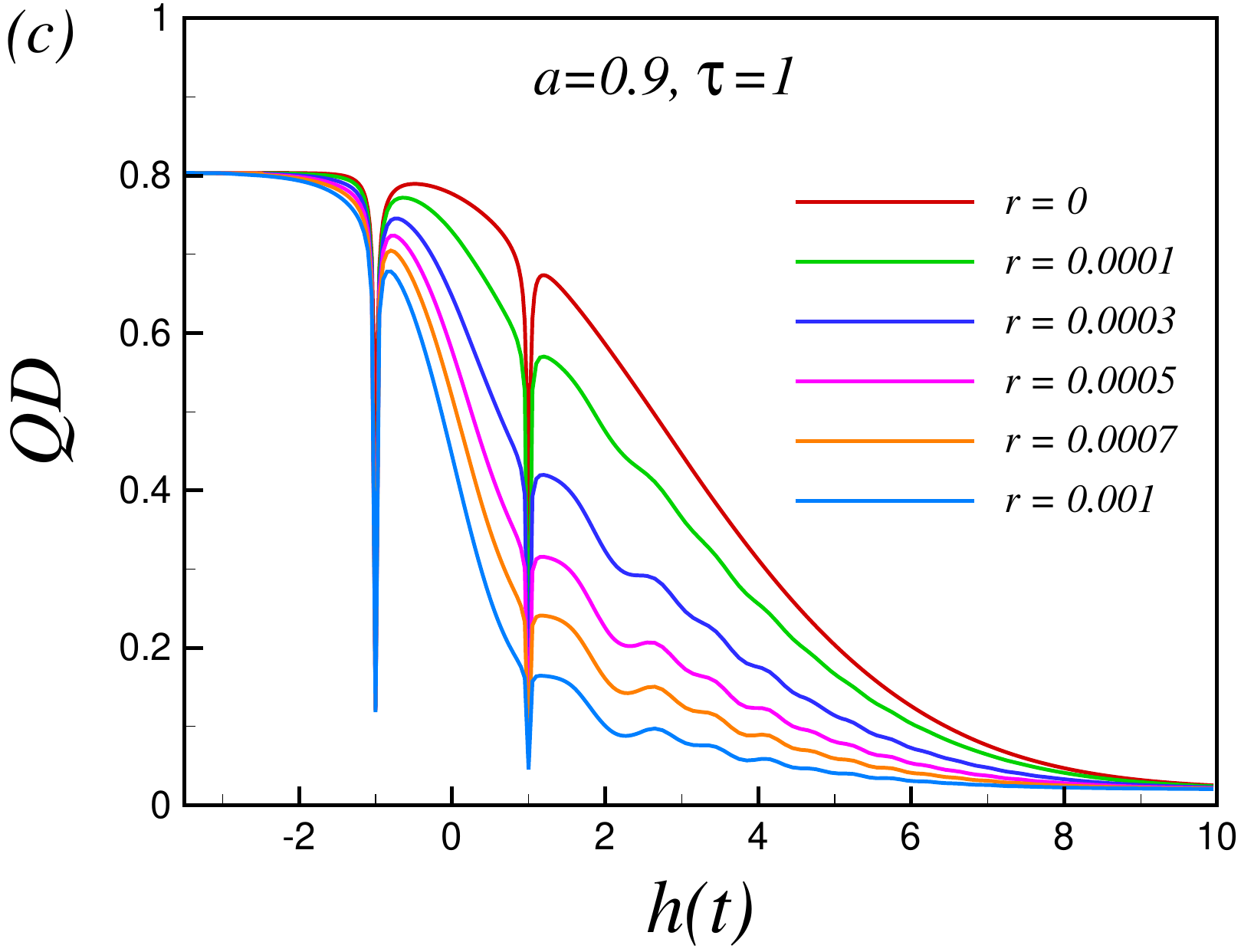}
\includegraphics[width=0.495\linewidth]{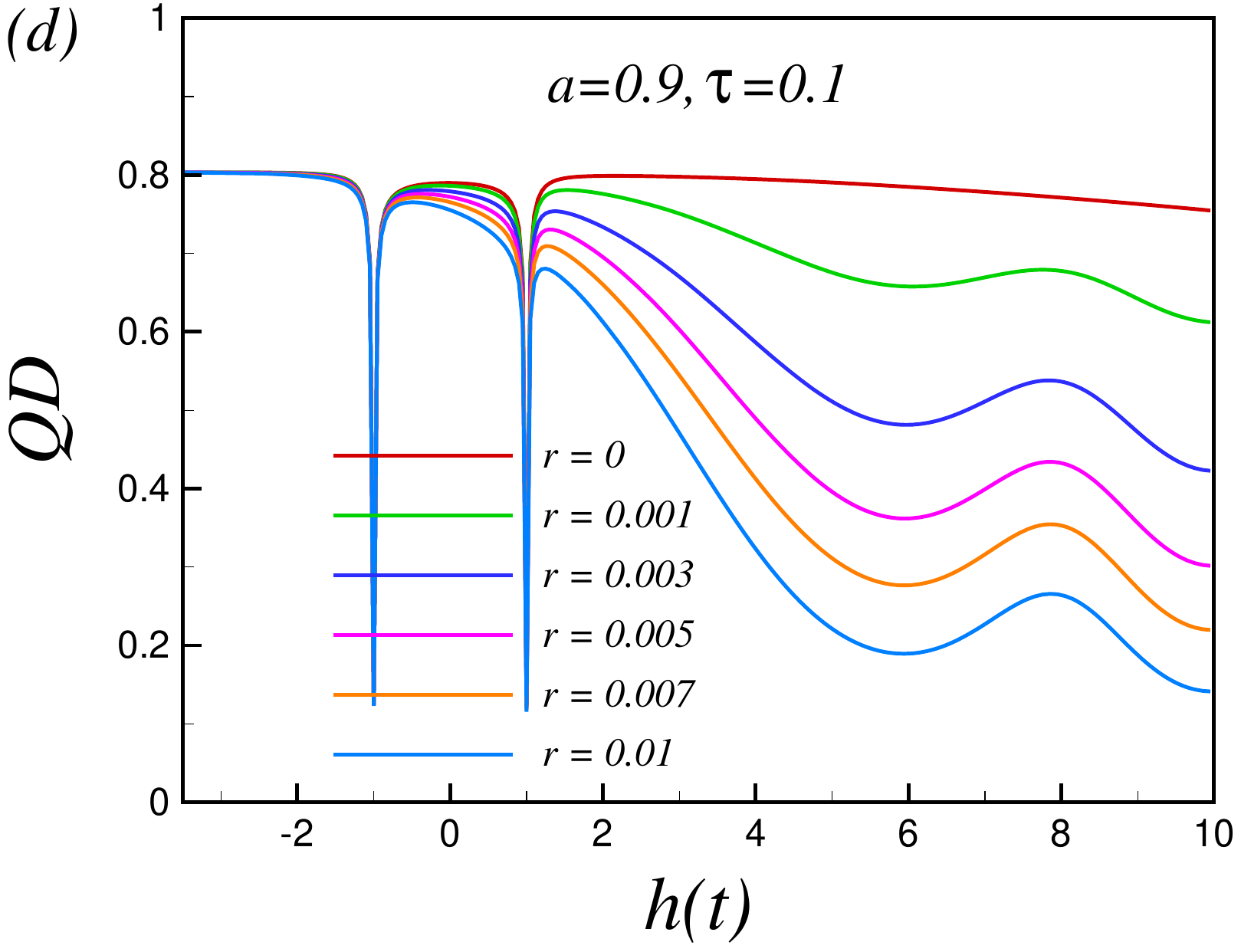}
\caption{Correlation dynamics under stochastic resetting for different ramp times, $N=500$, $\delta=0.01$, and $a=0.9$.  
(a,b) Concurrence $C$ as a function of reset rate $r$ for ramp times $\tau=1$ and $\tau=0.1$, respectively.  
(c,d) Quantum discord QD for the same ramp times.  
In both measures, resetting induces oscillatory suppression whose period grows as either the reset rate $r$ or the ramp time $\tau$ decreases.
}
\label{figAPB1}
\end{figure}

In the weak-coupling regime ($\tau=1$ and $\tau=0.1$), correlations decay monotonically and more rapidly under resetting, as shown in Figs.~\ref{fig4}(b)-(c) and \ref{fig5}(b)-(c).  
For shorter ramp times, the correlations are smaller in magnitude but also less sensitive to the reset dynamics.  
While in the reset-free case both concurrence and discord decay steadily after the transverse field crosses the second critical point, under resetting they display oscillatory suppression.  
The oscillation period increases as either the reset rate $r$ or the ramp time $\tau$ decreases (see Fig.~\ref{figAPB1} and  Appendix~\ref{APA}).  
Thus, even though the breakdown of adiabatic evolution at each critical point is still signaled by abrupt changes in concurrence and discord, the resetting process qualitatively reshapes their decay patterns.

\section{Summary and Conclusion}\label{sec:concl}
We  investigated the influence of stochastic quantum resetting on the dynamics of concurrence and quantum discord between two central qubits coupled to a driven transverse-field Ising chain environment.
The environment was linearly ramped across its quantum critical points, and resetting events projected it back to its initial state at random times.  
By extending the central spin/qubit framework to incorporate resetting, we showed that quantum resetting enhances the suppression of correlations even before the transverse field reaches the first critical point. This behavior contrasts with the reset-free case, where strong coupling and long ramp times lead to pronounced revivals of concurrence and discord between the critical points. Under resetting, these revivals are diminished: the peak heights of concurrence exhibit exponential decay with the reset rate, while discord shows no clear scaling. The origin of this difference can be traced to the structure of the decoherence factor, which accumulates multiplicatively across momentum modes and is highly sensitive to stochastic interruptions. Concurrence, being directly tied to phase-coherent overlaps, decays exponentially, while discord, which also includes classical contributions to correlations, is more robust and therefore shows no universal scaling. This complementary behavior highlights the importance of analyzing multiple correlation measures, as they capture different facets of quantum resources.  
In the weak-coupling regime, corresponding to short ramp times, correlations are  more strongly suppressed under resetting, though their sensitivity to the reset rate decreases. Furthermore, while in the absence of resetting both concurrence and discord decay monotonically as the system crosses the second critical point, resetting induces oscillatory decay patterns, with the oscillation period increasing as either the reset rate or the ramp time is reduced. We verified that these qualitative behaviors remain robust for different system sizes and for other initial states of the central qubits, indicating that the reported phenomena are not restricted to the Werner-state initialization.  
Overall, our results identify stochastic resetting as a versatile mechanism for reshaping correlation dynamics in many-body quantum systems. Beyond their fundamental significance, these findings suggest potential applications of resetting as a resource in quantum information processing, where controlled suppression or revival of correlations may be advantageous. Candidate experimental platforms include trapped-ion chains, Rydberg atom arrays, and superconducting qubits with engineered reset protocols, 
which offer a promising route to experimentally test the predicted revival and suppression dynamics.
It should be noted that our analysis was restricted to Poissonian reset protocols and one-dimensional environments; extending the framework to more general reset statistics and higher-dimensional systems remains an open challenge.

%
%

\section{Acknowledgement}
This work is based upon research funded by Iran National Science Foundation (INSF) under project No. 4024646.
A.~Akbari is supported by the Beijing Natural Science Foundation (Grant No. IS25015).

\appendix
\section{Analytical Framework for the Driven Ising Environment \label{APA}} 

The Hamiltonian of the Ising chain with periodic boundary conditions and subject to a time-dependent transverse magnetic field is  
\begin{equation}
\label{eq:Ising}
{\cal H}(t) = -\sum_{j=1}^{N} \big(\sigma_j^x \sigma_{j+1}^x + h(t)\sigma_j^z \big),
\end{equation}
where $\sigma^{x,z}_j$ are Pauli matrices acting on site $j$ of a one-dimensional lattice. For a static field $h(t)=h$, the ground state is ferromagnetic with $\langle \sigma_n^x \rangle \neq 0$ for $|h|<1$, and paramagnetic with $\langle \sigma_n^x \rangle =0$ otherwise. These phases are separated by quantum critical points at $h=\pm 1$~\cite{Pfeuty1970}.  

Using a Jordan-Wigner transformation~\cite{LSM1961}, ${\cal H}(t)$ can be mapped onto a model of spinless fermions,  
\begin{equation}
\label{eq:FF}
\bl
{\cal H}(t) =
&
 \sum_{k>0} \Big[-\big(h(t)-\cos k\big)\,(c_{k}^{\dagger}c_{k}+c_{-k}^{\dagger}c_{-k})
\\
&
+ \sin k \,(c_{k}^{\dagger}c_{-k}^{\dagger}+c_{k}c_{-k}) \Big],
\el
\end{equation}
with fermionic anihilation (creation) perators $c_n$~($c_n^\dagger$). A Fourier transformation,  
$c_n = (e^{i\pi/4}/\sqrt{N}) \sum_k e^{ikn}c_k$ (the phase factor $e^{i\pi/4}$ is introduced for convenience),  
leads to a decomposition ${\cal H}(t)=\sum_{k>0} C_k^\dagger {\cal H}_k(t) C_k$,  
where $C_k^\dagger=(c_k^\dagger,\;c_{-k})$ is a Nambu spinor and  
${\cal H}_k(t) = h_k(t)\sigma^z + \Delta_k\sigma^x$, with $h_k(t)=-(h(t)-\cos k)$ and $\Delta_k=\sin k$.  

For a linear ramp $h(t)=t/\tau$ and defining $|\varphi_{k}(t)\rangle=(v_k(t),u_k(t))^T$, the amplitudes satisfy the time-dependent Schrödinger equations~\cite{Vitanov1999,Damski2011,Suzuki2016,Nag2012}  
\begin{eqnarray}
\label{eq8}
\bl
i \frac{d}{dt} v_{k}(t) &= -(h(t)-\cos k)\,v_{k}(t) + \sin k \, u_{k}(t), \\
i \frac{d}{dt} u_{k}(t) &= \phantom{-}(h(t)-\cos k)\,u_{k}(t) + \sin k \, v_{k}(t). \quad
\el
\end{eqnarray}
The solutions can be written as
$v_k(t)=U_{11}(t)v_{k}(0)+U_{12}(t)u_{k}(0)$ and $u_k(t)=U_{21}(t)v_{k}(0)+U_{22}(t)u_{k}(0)$,  
with matrix elements  
%
\begin{widetext}
\begin{eqnarray}
\bl
\no
U_{11}(t) =& \frac{\Gamma(1-\omega)}{\sqrt{2\pi}} 
\Big[D_{\omega-1}(-z_i)D_{\omega}(z_f)
+D_{\omega-1}(z_i)D_{\omega}(-z_f)\Big], 
\\
U_{12}(t) =& \frac{\Gamma(1-\omega)}{\lambda\sqrt{\pi}}\,e^{i\pi/4}
\Big[D_{\omega}(z_i)D_{\omega}(-z_f)
-D_{\omega}(-z_i)D_{\omega}(z_f)\Big], 
\\
U_{21}(t) =& \frac{\lambda\Gamma(1-\omega)}{2\sqrt{\pi}}\,e^{-i\pi/4}
\Big[D_{\omega-1}(z_i)D_{\omega-1}(-z_f)
-D_{\omega-1}(-z_i)D_{\omega-1}(z_f)\Big], 
\\
U_{22}(t) =& \frac{\Gamma(1-\omega)}{\sqrt{2\pi}} 
\Big[D_{\omega}(-z_i)D_{\omega-1}(z_f)+D_{\omega}(z_i)D_{\omega-1}(-z_f)\Big], 
\el
\end{eqnarray}
%
where $\Gamma(\nu)$ is the Euler gamma function, $D_\omega(z)$ denotes the parabolic cylinder function~\cite{szego1954,abramowitz1988}, $\omega=i\lambda^2/2$, $\lambda=\Delta_k\sqrt{\tau}$, $z_i=\sqrt{2}\,e^{-i\pi/4}\tau_k(t_i)/\sqrt{\tau}$, and $\tau_k(t)=h_k(t)\tau$.  
Note: for effective fields $h_k(t)=-(h(t)\pm\delta-\cos k)$, the wavefunction is defined as $|\psi_k^\pm(t)\rangle=(v_k^\pm(t),u_k^\pm(t))^T$.
\end{widetext}

\bibliography{REFQRD}

\end{document}